\documentstyle[12pt,epsf]{article} 

\renewcommand{\thefootnote}{\fnsymbol{footnote}}
\renewcommand{\theequation}{\arabic{section}.\arabic{equation}}

\begin{document}


\begin{titlepage}
\begin{flushright}
\begin{tabular}{l}
FERMILAB--PUB--96/455--T\\
NORDITA--96--81--P\\
hep-ph/9701238
\end{tabular}
\end{flushright}
\vskip0.5cm
\begin{center}
  {\Large \bf Use and Misuse of QCD Sum Rules in Heavy-to-light
  Transitions: the Decay $B\to\rho e \nu$ Reexamined\\}
  \vskip1cm {\large Patricia Ball}
  \vskip0.2cm
  Fermi National Accelerator Laboratory,
  P.O.\ Box 500, Batavia, IL 60510, USA\\
  \vskip0.5cm {\large V.M.\ Braun}\footnote{On leave of absence from
    St.\ Petersburg Nuclear Physics Institute, 188350 Gatchina, Russia.}\\
  \vskip0.2cm
  NORDITA, Blegdamsvej 17, DK--2100 Copenhagen, Denmark\\
  \vskip1.8cm
  {\large\bf Abstract:\\[10pt]} \parbox[t]{\textwidth}{ 
The existing calculations of the form factors describing the decay
    $B\to\rho e \nu$ from QCD sum rules have yielded conflicting
    results at small values of the invariant mass
    squared of the lepton pair.  We demonstrate that the
    disagreement originates from the failure of the short-distance
    expansion to describe the $\rho$ meson distribution amplitude in
    the region where almost the whole momentum is carried by one of
    the constituents. This limits the applicability of QCD sum rules
    based on the short-distance expansion of a three-point correlation
    function to heavy-to-light transitions and calls for an expansion
  around the light-cone, as realized in the light-cone sum rule
  approach. We derive and update light-cone sum rules for all the
  semileptonic form factors, using recent results on the $\rho$ meson
  distribution amplitudes. 
  The results are presented in detail
  together with a careful analysis of the uncertainties, 
  including estimates of higher-twist effects, and compared
  to lattice calculations and recent CLEO measurements. We also derive
  a set of
    ``improved'' three-point sum rules, in
    which some of the problems 
  of the  short-distance expansion are avoided and whose results 
  agree to good accuracy with those from light-cone sum rules.}
  \vskip1cm 
{\em Submitted to Physical Review D}
\end{center}
\end{titlepage}

\renewcommand{\thefootnote}{\arabic{footnote}}
\setcounter{footnote}{0}


\section{Introduction}
\setcounter{equation}{0}

The interest in the study of semileptonic $B$ decays is mainly due to
their importance in determining the CKM matrix elements $|V_{cb}|$ and
$|V_{ub}|$. Whereas the theoretical analysis of both exclusive and
inclusive $b\to c$ transitions
is decisively simplified by an expansion in the inverse heavy quark
mass, this method is of only little use in $b\to u$
transitions. This is essentially due to the fact that in inclusive
channels 
experimental observation is possible only in a small region of
phase-space beyond the kinematical threshold for charm production, in which
the hadron multiplicity is small. Thus, since the theoretical
description of inclusive
decays is essentially based on a parton-model picture, it is not
very predictive in the experimentally accessible
range, cf.~\cite{buincl}. It is therefore rather the exclusive decay 
channels, in 
particular $B\to\pi\ell\nu$ and $B\to\rho\ell\nu$, that seem to be 
more suitable for obtaining reliable information on
$|V_{ub}|$. The CLEO collaboration has recently presented first
experimental results of these branching ratios \cite{CLEO}, which,
however, are model-dependent.

The decay $B\to\pi\ell\nu$ has been tackled by several authors
using a number of different approaches, in particular QCD sum rules
\cite{BPi,narrdecay,PRD48,BKR93} and simulations on the lattice 
\cite{ELC,APE,lattBPi,lellouch}; the results are in reasonable
agreement. The
situation is, however, not that favourable in the $B\to\rho$
channel. Although here the same methods were applied, the resulting
predictions for the decay rates are quite different. To illustrate the
origin of the
problem, let us first introduce the relevant observables:
the hadronic matrix element determining the $B\to\rho$ weak transition
is 
\begin{eqnarray}
\langle \rho,\lambda | (V-A)_\mu | B \rangle & = &
-i (m_B + m_\rho) A_1(t) \epsilon_\mu^{*(\lambda)} +
\frac{iA_2(t)}{m_B
+ m_\rho} (\epsilon^{*(\lambda)}p_B) (p_B+p_\rho)_\mu\nonumber\\
& & {} + \frac{iA_3(t)}{m_B + m_\rho} (\epsilon^{*(\lambda)}p_B)
(p_B-p_\rho)_\mu + \frac{2V(t)}{m_B + m_\rho}
\epsilon_\mu^{\phantom{\mu}\alpha\beta\gamma}\epsilon_\alpha^{*(\lambda)}
p_{B\beta} p_{\rho\gamma},\makebox[0.8cm]{}\label{eq:ME}
\end{eqnarray}
where the four form factors $A_{1,2,3}$ and $V$ depend on the momentum
transfer $t=(p_B-p_\rho)^2$ to the leptons; in the limit of vanishing
lepton mass $A_3$ does not contribute to the semileptonic decay rate
and will not be considered in this paper. $\lambda$ is the
polarization of the $\rho$, $t$ takes values between 0 and
20.3~GeV$^2$. It is precisely this
large range of relevant $t$ that renders the theoretical description
of the form factors in (\ref{eq:ME}) so difficult. Most quark model 
calculations rely
essentially on the pole-dominance picture \cite{BWS} or on a
nonrelativistic description, which yields an exponential increase of
the form factors with $t$ \cite{ISGW}, which was softened
in an updated version of the model, Ref.~\cite{ISGWrev}. 
Lattice calculations are up to
now limited to small momenta of the final state $\rho$
\cite{ELC,APE,UKQCD} and/or require extrapolations in the heavy quark 
mass. For $B\to
\pi\ell\nu$, the possibility to restrict the functional
$t$ dependence of the single relevant form factor from unitarity
with the supplementary input of available lattice data at large $t$
was investigated in \cite{lellouch}, but this method is presently not
very predictive in the $\rho$ channel, see \cite{rho_disprel}.

To date, only lattice simulations and QCD sum rules seem to be apt to {\em
predict} the $t$ dependence in nearly the whole physical region.
QCD sum rules provide a consistent description of semileptonic $D$ decays
\cite{BBD} and  of the $B\to\pi$ semileptonic form factor 
\cite{BPi,narrdecay,PRD48,BKR93}. However, there exist conflicting 
predictions from different types of QCD sum rules 
for $B\to\rho$ decays, which differ by a factor~2 in the form factors 
at maximum recoil \cite{PRD48,ABS}. The 
aim of this paper is to clarify the origin of this difference and to
give updated predictions for the form factors, which include in
particular recently gained information on the structure of $\rho$ mesons
probed at large momentum transfer \cite{rhoWF}.

\begin{figure}
\vspace*{-2.4cm}
\centerline{\epsfxsize12.0cm\epsffile{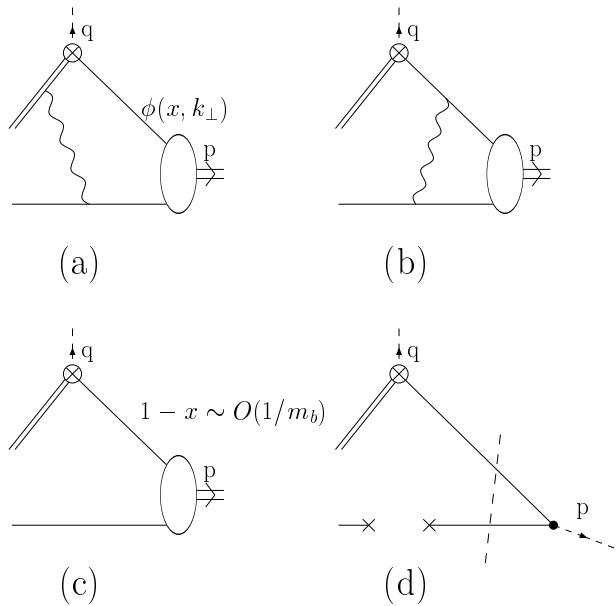}}
\vspace*{-6.5cm}
\caption[]{``Hard'' (a,b) and ``soft'' (c) contributions to the decay 
form factor. (d): modelling of the soft contribution in the QCD sum rule
approach. See explanation in the text.}
\end{figure}
At large recoil, the light quark originating
from the weak decay carries a large energy of order $m_b/2$ and
has to transfer it to the soft cloud to recombine to the final 
state hadron. 
The probability of such a recombination depends on the parton content
of both the $B$ meson and the light meson, the valence configuration with
the minimum number of Fock constituents 
being dominant. 
The valence quark configuration is characterized by the wave function
$\phi(x,k_\perp)$ depending on the momentum fraction $x$ carried by the
quark and on its transverse momentum $k_\perp$. There exist two different 
mechanisms for the valence quark contribution to the transition form 
factor.
The first one is the hard rescattering mechanism pictured in Figs.~1(a,b), 
which requires that
the recoiling and spectator quarks are at small transverse separations.
In this case the large momentum is transferred by an exchange
of a hard gluon with virtuality $k^2\sim O(m_b)$. This contribution is 
perturbatively calculable in terms of the 
Bethe-Salpeter wave functions at 
small ($\sim 1/m_b$) transverse separations, or 
{\em distribution amplitudes}:
\begin{equation}
\phi(x) =\int^{k_\perp^2\sim m_b}dk_\perp^2
\phi(x,k_\perp).
\end{equation}
The second mechanism is the soft contribution, shown schematically in 
Fig.~1(c).
The idea is that hard gluon exchange is not necessary, provided 
one picks up an ``end-point'' configuration with almost all 
momentum $1-x\sim O(1/m_b)$ carried by one constituent.
Since at large scales  
$\phi(x)\sim 1-x$ \cite{exclusive}, the overlap integral 
is of order $\int_{1-O(1/m_b)}^1 dx\, \phi(x) \sim O(1/m_b^2)$.
An additional factor $m_b^{1/2}$ comes from the normalization of
the heavy initial state, so that the final scaling law for the soft 
contribution to the form factor at large recoil is 
$1/m^{3/2}_b$ \cite{CZ90,ABS,ASY94}.
Note that the transverse quark-antiquark separation is not constrained
in this case, which implies that the soft contribution is sensitive 
to long-distance dynamics. To calculate the soft
contribution one needs to know the wave function as a function of the
transverse separation; the simpler distribution amplitude is not enough.

Hard exclusive processes involving light hadrons receive the same 
two types of contributions. There is a major difference, however,
in that for light hadrons the soft contribution is suppressed by a power
of the large momentum (i.e.\ it is of higher-twist), while for heavy
meson decays both soft and hard contributions turn out to be of the 
same power in the heavy quark mass \cite{CZ90,ABS,ASY94}. 
As a result, the soft (end-point) contribution is expected to 
be large and requires quantitative evaluation.
  
It was suggested \cite{ASY94} that Sudakov-type perturbative corrections
cut off contributions of large transverse separations so that
the soft contributions might be suppressed. This suppression
eventually eliminates the soft contribution for very large $m_b$.  
At the realistic values $m_b\approx 5\,$GeV,
however, it is unlikely that  calculations of this type
can provide a quantitative
description. Indeed, the existing estimates of ``hard'' contributions
typically fall short of realistic values of the form factors from
model calculations.\footnote{ 
Note that a
similar suppression is present for hard exclusive processes involving
light hadrons \cite{LS92} (in which case the soft contribution is 
additionally 
suppressed by a power of the large momentum); however, there is increasing
evidence that soft contributions to, say, the pion form factor
remain important at least up to $Q^2\approx 10\,$GeV$^2$, see 
\cite{softpion}.}

Here the QCD sum rules method enters the stage and suggests a 
nonperturbative
technique to estimate the necessary convolution integral without
explicit knowledge of the wave functions.

There exist two different types of QCD sum rules, which, being similar
in spirit, differ in the treatment of the light hadron in the final
state. This is illustrated in Figs.~1(c,d).

The ``traditional'' sum rules avoid introducing wave functions
altogether by
considering a three-point correlation function with a suitable
interpolating current and use dispersion relations to extract the 
contribution of the ground state. The most important nonperturbative effect
is then described by the diagram in Fig.~1(d), where the light quark
is soft and interacts with the nonperturbative QCD vacuum, forming the
so-called quark condensate. Since quarks in a condensate
have zero momentum, it is clear that this diagram yields a
contribution to the distribution amplitude that
is naively proportional to $\delta(1-x)$ and remains
unsuppressed for $m_b\to\infty$.
This obviously violates the power counting discussed above.
The contradiction must be resolved by including the contributions of
higher-order condensates to the sum rules and subtracting the 
contribution of excited states.
The suppression of the end-point region $x\to 1$, which is {\em strictly}
required by QCD, is thus expected to hold as a {\em numerical}
cancellation between different contributions, which becomes the more
delicate (and requires more fine-tuning) the more $m_b$ increases.
For $m_b\approx 5\,$GeV a suppression of
the quark condensate contribution by a factor
$\sim 1\,$GeV$^2/m_b^2\sim 1/25$ is required. 
This explains why the traditional three-point sum rules
become unreliable.

The light-cone sum rules avoid this problem by rearranging the calculation
in such a way that nonperturbative effects like the
interaction with the quark condensate are included 
in the {\it nonperturbative} hadron distribution amplitudes, 
estimated using additional sum rules \cite{CZreport}.
These additional sum rules are written 
for integrated characteristics of the distribution amplitudes like moments,
and the correct asymptotic behaviour at the end-points is included
by construction.

The contribution of a single leading-twist distribution amplitude 
incorporates
an infinite set of contributions of condensates of increasing dimension
in the standard approach, at the cost of retaining pieces  
with the largest Lorentz spin (lowest twist) only.
The remaining pieces of condensate contributions are organized in a similar
way in the contributions of higher-twist distribution amplitudes.\footnote{
While the usual sum rules are based on matching the QCD
calculation at small distances to the phenomenological description in
 terms of hadrons at large distances, the light-cone sum rules 
 match in {\em transverse distance}. Hence the relevant 
parameter in the expansion is twist, and not dimension.} 

The premium for this rearrangement
is that light-cone sum rules  have the correct 
asymptotic behaviour in the heavy quark limit, but the snag is that
the present knowledge of higher-twist distribution amplitudes is 
incomplete, so
that not all known nonperturbative corrections of the standard approach 
can be included. One should expect that
these two approaches provide complimentary descriptions of $B$ decays,
with their own advantages and disadvantages.

Note that the problem with three-point sum rules originates from the  
constraint of the distribution amplitude convolution integral to the 
end-point region;
this makes the answer very sensitive to the precise {\em shape} 
of the distribution amplitude rather than its integrated characteristics.
For the decay form factors at small recoil there is no such restriction,
and the classical QCD sum rule concept of taking into account
nonperturbative effects as the contributions of  long-wave vacuum 
fluctuations (condensates) classified by their dimension 
is adequate. Thus, at small recoil,  both types of the sum rules 
should yield similar results
and the spread of their predictions characterizes the accuracy of the 
method.
At large recoil one should rely on the light-cone sum rules.   

It is worthwhile to add that contributions of hard rescattering 
can be consistently included as radiative corrections to the sum rules.
Their inclusion is technically challenging, but does not pose a 
problem of principle.

Our paper is organized as follows. In Sec.~2 we introduce
the two different types of QCD sum rules. In Sec.~3 we then analyse
and explain the discrepancy in $B\to\rho$ transitions, paying
special attention to the rearrangement of quark and mixed condensate
contributions within the light-cone expansion. We also estimate
higher-twist contributions to $A_1$. In Sec.~4 we discuss shortly
the quark mass dependence and the heavy quark limit.  Section~5 
contains our final predictions for form factors, spectra and decay
rates of $B\to\rho e\nu$, Sec.~6 a summary and the discussion of the
results. More technical issues are delegated to appendices.


\section{Three-Point vs.\ Light-Cone Sum Rules}
\setcounter{equation}{0}

In this section we discuss the two different types of QCD sum rules
 that have been used in the literature to calculate heavy-to-light
 meson decays. 

\subsection{Three-Point Sum Rules}

The -- comparatively speaking -- ``traditional'' approach towards
transition matrix elements is by calculating a three-point (``3pt'' in
the following) correlation function. Specifically, 
for $B\to\rho$ one considers\footnote{For
  3pt functions we follow the notations of \cite{PRD48}.}
\begin{eqnarray}
\Gamma^{\mu\nu} & = & i^2\!\!\int d^4x\, d^4y\, e^{-ip_B x + i p_\rho
 y}\, \langle 0 | T j_\rho^\nu(y) (V-A)^\mu(0) j_B^\dagger (x) | 0
\rangle\nonumber\\
& = & i g^{\mu\nu} \Gamma_0 - i (p_B+p_\rho)^\mu p_B^\nu \Gamma_+ -
\epsilon^{\mu\nu}_{\phantom{\mu\nu}\rho\sigma} p_B^\rho p_\rho^\sigma
\Gamma_V + \dots\label{eq:corrga}
\end{eqnarray}
Here $(V-A)_\mu = \bar u \gamma_\mu(1-\gamma_5)b$ is the weak current
mediating the $b\to u$ transition, $j_B = \bar q i\gamma_5 b$ is the
interpolating field for the pseudoscalar $B$ meson, and $j_\rho^\nu =
\bar q \gamma^\nu u$ the interpolating field for the $\rho$ meson. In
(\ref{eq:corrga}) we have given explicitly only those Lorentz
structures that are relevant for the semileptonic decay channel, the others
are suppressed. 

The method of QCD sum rules consists -- in principle -- in performing
on the one hand
a QCD calculation of $\Gamma^{\mu\nu}$ in the not
so deep Euclidean region $p_\rho^2 \sim p_B^2-m_b^2 \sim
O(1\,$GeV$^2$), writing it on the other hand as (double)
dispersion relation over the physical cut and equating both
expressions. It was the idea of Shifman, Vainshtein and Zakharov
\cite{SVZ} to complement the purely perturbative calculation of
$\Gamma^{\mu\nu}$ by nonperturbative terms in form of vacuum matrix 
elements of gauge-invariant local operators, the condensates.
The method proved surprisingly successful in describing various 
hadronic matrix elements in terms of a handful of input parameters, as
is testified by the immense number of publications in the field.

The traditional approach by SVZ appeals to Wilson's 
operator product expansion (OPE), which is the expansion of a
T-product of currents at short distances in terms of
local operators. 
In that way  one obtains for the invariants $\Gamma$ in (\ref{eq:corrga}):
\begin{equation}\label{eq:OPE}
\Gamma(p_B^2,p_\rho^2,t) = \sum_n \Gamma^{(n)}(p_B^2,p_\rho^2,t)
\langle\, 0\,|\, {\cal O}_n\,| \, 0\, \rangle
\end{equation}
with the condensates $\langle\, 0\,|\, {\cal O}_n\,| \, 0\, \rangle$.
In most applications one takes into account condensates up to
dimension 6. $\Gamma$ can also be expressed as a dispersion relation
over the physical singularities: 
\begin{equation}
\Gamma = \int\!\! ds_B\,ds_\rho\,
\frac{\rho^{\rm phys}(s_B,s_\rho,t)}{(s_B-p_B^2)(s_\rho-p_\rho^2)} + {\rm
  subtractions}.
\label{eq:doubleDD}
\end{equation}
Usually one is interested only in the properties of the ground state,
which has to be extracted from the sum over all states. For this, one 
writes
\begin{equation}
\rho^{\rm phys} = \rho^{\rm ground\ state} + \rho^{\rm cont}
\end{equation}
and approximates the unknown contribution of the continuum by the
perturbative spectral function above some ``continuum-thresholds''
$s_0^B$ and $s_0^\rho$, such that
\begin{equation}
\rho^{\rm cont} \simeq \rho^{\rm pert}
\{1-\Theta(s_B^0-s_B)\, \Theta(s_\rho^0-s_\rho)\},
\end{equation}
where the ``$\simeq$'' indicates that smearing over a sufficiently
large interval is implied. The sum rule is then obtained by equating
(\ref{eq:OPE}) and (\ref{eq:doubleDD}) and subtracting from both sides
the continuum contribution, i.e.\ in the above approximation the 
integral over the perturbative double spectral function above
thresholds. In order to
reduce the impact of this approximation on the final results as well as
the error induced by truncating the series (\ref{eq:OPE}) after the
first few terms, one subjects the sum rule to a Borel
transformation. For an arbitrary function of Euclidean momentum,
$f(P^2)$ with $P^2=-p^2$, the transformation is defined by:
\begin{equation}\label{eq:defborel}
\hat{f} := \widehat B_{P^2}(M^2)\, f = \lim_{\begin{array}{c}
\scriptstyle P^2\to\infty,N\to\infty\\[-1mm]
\scriptstyle P^2/N = M^2{\rm\ fixed} \end{array}}
\frac{1}{N!} (-P^2)^{N+1} \frac{d^{N+1}}{(dP^2)^{N+1}}\,f,
\end{equation}
where $M^2$ is the so-called Borel parameter. For a typical term in
the OPE, the transformation yields
\begin{equation}
\widehat B_{P^2}(M^2) \,\frac{1}{(p^2-m^2)^n} =
\frac{1}{(n-1)!}\, (-1)^n\, \frac{1}{(M^2)^n}
e^{-m^2/M^2}.
\end{equation}
As condensates with large dimension get divided by correspondingly
high powers of $(p^2-m^2)$, their contributions to the sum rules get
suppressed by factorials. In the dispersion integrals,
$\rho^{\rm cont}$ gets exponentially suppressed relatively
to the ground state contribution, which is just the desired effect.

Defining the couplings of the mesons to their interpolating fields by
\begin{equation}
\langle \, 0 | \, \bar d i \gamma_5 b \, | \, \bar B^0 \, \rangle 
=  f_B \frac{m_B^2}{m_b}\,,\qquad
\langle \, 0 \, | \, \bar d \gamma_\nu u \, | \, \rho^+, \lambda\,
\rangle  =  f_\rho m_\rho \epsilon^{(\lambda)}_\nu,
\end{equation}
we find the following sum rules for the form factors determining the
semileptonic $B\to\rho$ transition:
\begin{eqnarray}
A_1^{B\to\rho}(t) & = & \frac{m_b}{f_Bf_\rho (m_B+m_\rho) m_B^2 m_\rho}\,
\exp \left\{ \frac{m_B^2}{M_B^2} + \frac{m_\rho^2}{M_\rho^2} \right\}
M_B^2 M_\rho^2 \hat{\Gamma}_0,\label{eq:3ptA1}\\
A_2^{B\to\rho}(t) & = & \frac{m_b(m_B+m_\rho)}{f_Bf_\rho m_B^2m_\rho}\,
\exp \left\{ \frac{m_B^2}{M_B^2} + \frac{m_\rho^2}{M_\rho^2} \right\}
M_B^2 M_\rho^2 \hat{\Gamma}_+,\\
V^{B\to\rho}(t) & = & \frac{m_b(m_B+m_\rho)}{2f_Bf_\rho m_B^2m_\rho}\,
\exp \left\{ \frac{m_B^2}{M_B^2} + \frac{m_\rho^2}{M_\rho^2} \right\}
M_B^2 M_\rho^2 \hat{\Gamma}_V,\label{eq:SR3pt}
\end{eqnarray}
where the $\hat\Gamma$ on the right-hand sides denote the
Borel-transformed invariants after continuum subtraction. The explicit
formulas for $\hat{\Gamma}$ can be found in \cite{BBD,PRD48}. 

\subsection{Light-Cone Sum Rules}

An alternative approach \cite{BBK,BF1,CZ90}  
starts from the two-point function sandwiched between the vacuum
and the $\rho$ meson state:
\begin{eqnarray}
\Pi_\mu & = & i\!\!\int d^4x\, e^{-ip_Bx}\, 
\langle \rho(p_\rho,\lambda) | 
T (V-A)_\mu(0) j_B^\dagger (x) | 0 \rangle\nonumber\\
& = &
-i \Pi_1(p^2_B,t) \epsilon_\mu^{*(\lambda)} +
i\Pi_2(p^2_B,t) (\epsilon^{*(\lambda)}p_B) (p_B+p_\rho)_\mu
+ \Pi_V(p_B^2,t)
\epsilon_\mu^{\phantom{\mu}\nu\alpha\beta}\epsilon_\nu^{*(\lambda)}
p_{B\alpha} p_{\rho\beta}+\dots,\label{eq:cor2pt}
\end{eqnarray}
with $p_B^2-m_b^2 < 0$ and $p_\rho^2=m_\rho^2$.
Hence we encounter only single dispersion relations,
\begin{equation}\label{eq:2.11}
\Pi = \int\!\! ds_B\,
\frac{\rho^{\rm phys}(s_B,t)}{s_B-p_B^2} + {\rm
  subtractions},
\end{equation}
and to isolate the ground state only an approximation
 for the continuum contribution in the $B$ meson channel is needed:
\begin{equation}
\rho^{\rm cont} \simeq \rho^{\rm pert}
\{1-\Theta(s_0^B-s_B)\}.
\end{equation}
Thus, this part becomes simpler and potentially more accurate than
with 3pt sum rules, since less assumptions are made. 

The price to pay, however, is that the
QCD calculation becomes more involved. In particular, the expansion
of (\ref{eq:2.11}) in local operators becomes useless since 
an infinite sequence of such operators contributes to the same order in 
$1/(p_B^2-m_b^2)$.
Indeed, each operator of the sequence
$$\bar q D_{\mu_1}\cdots D_{\mu_n} \Gamma q, $$
where $D_{\mu_i}$ is the covariant derivative and
$\Gamma$ an arbitrary Dirac matrix structure,
symmetrized over the indices $\mu_1,\dots, \mu_n$ and with subtracted 
traces,
enters with the same suppression factor $1/(p_B^2-m_b^2)$ \cite{Bel95}. 
This is different from 3pt sum rules where contributions 
with higher $n$ are supressed by powers of $1/p_\rho^2$, which here
is no longer an expansion parameter. 
Still, contributions of operators containing {\em traces} over Lorentz
indices, or transverse components of the gluon fields are suppressed 
by extra powers of  $1/(p_B^2-m_b^2)$. This means that the relevant 
parameter is the operator {\em twist} rather than  dimension. 
The expansion goes in terms of {\em nonlocal} string-like operators on
the light-cone (``LC'' in the following), whose 
vacuum-to-meson transition amplitudes define the meson LC
distribution amplitudes, which describe the momentum fraction distribution 
among the meson constituents. The leading-twist distributions
correspond to the minimum number of Fock constituents and in the
case of a charged $\rho$ meson involve the following functions 
\cite{ABS,rhoWF}:
\begin{eqnarray}
\langle \rho^+(p,\lambda) |\bar u(0)\sigma_{\mu\nu}d(x)
|0\rangle & = &
-if_\rho^\perp(\mu)(e^{*(\lambda)}_\mu p_\nu -e^{*(\lambda)}_\nu p_\mu)
 \int_0^1 du\, e^{iupx} \phi_\perp(u,\mu),
\label{def1}\\
\langle \rho^+(p,\lambda) |\bar u(0)\gamma_\mu d(x)
|0\rangle &=& p_\mu \frac{(e^{*(\lambda)} x)}{(px)}
f_\rho m_\rho\int_0^1 du\, e^{iupx} \phi_\parallel(u,\mu)
\nonumber\\
&&\mbox{}+\left( e^{*(\lambda)}_\mu -p_\mu \frac{(e^{*(\lambda)} x)}{(px)}
\right) f_\rho m_\rho
\int_0^1 du\, e^{iupx} g_\perp^{(v)}(u,\mu)\,,
\label{def2}\\
\langle \rho^+(p,\lambda) |\bar u(0)\gamma_\mu\gamma_5 d(x)
|0\rangle &=&
\frac{1}{4} \epsilon_{\mu\nu\rho\sigma} e^{*(\lambda) \nu}
p^\rho x^\sigma  f_\rho m_\rho
\int_0^1 du\, e^{iupx} g_\perp^{(a)}(u,\mu)\,.
\label{def3}
\end{eqnarray}
For the sum rules for the $B\to\rho$ form factors, we also need the
function 
\begin{equation}
\Phi_\parallel(u,\mu) = \frac{1}{2}\,\left[ \bar u \int_0^u\!\!
dv\,\frac{\phi_\parallel(v,\mu)}{\bar v} - u \int_u^1\!\!
dv\,\frac{\phi_\parallel(v,\mu)}{v}\right].\label{eq:bigphi}
\end{equation}
In the above definitions the matrix elements are not gauge-invariant,
but refer to the axial gauge $x_\mu A^\mu(x) = 0$. In a general gauge,
gauge factors
\begin{equation}
\mbox{Pexp}\left[ig\int_0^1 d\alpha\, x^\mu A_\mu(\alpha
x)\right]\label{eq:gauge-factor}
\end{equation}
have to be inserted in between the quark fields.
The integration variable $u$ corresponds to the
momentum fraction carried by the quark.  The normalization is such that
$$\int_0^1 du\, f(u)=1$$
for all four distributions $f=
\phi_\perp , \phi_\parallel , g_\perp^{(v)} ,
g_\perp^{(a)}$. $f_\rho^\perp(\mu)$, the scale-dependent 
coupling of the $\rho$ meson
to the tensor current, is defined by (\ref{def1}) for $x=0$.
To leading-twist 2 accuracy the ``$g$-functions'' are in fact not
independent, but related to $\phi_\parallel$ by
 Wandzura-Wilczek \cite{WW} type relations:
\begin{eqnarray}
g_\perp^{(v),{{\rm twist}\, 2}}(u,\mu) &=&
\frac{1}{2}\left[
\int_0^u dv \frac{\phi_\parallel(v,\mu)}{\bar v}
                 + \int_u^1 dv \frac{\phi_\parallel(v,\mu)}{ v}\right]\,,
\nonumber\\
    g_\perp^{(a),{{\rm twist}\, 2}}(u,\mu) &=&
2\left[
 \bar u \int_0^u dv \frac{\phi_\parallel(v,\mu)}{\bar v}
                 + u \int_u^1 dv \frac{\phi_\parallel(v,\mu)}{ v}
\right] \,.
\label{WW1}
\end{eqnarray}
All these functions are discussed in detail in \cite{rhoWF}.
The distribution amplitudes of higher-twist can be related to
contributions of Fock-states with more constituents (say, an extra gluon)
and generally generate contributions to the sum rules that are suppressed
by powers of $1/(p_B^2-m_b^2)$. We will discuss higher-twist
distribution amplitudes only shortly in Sec.~3.

Note that the matrix element of a single nonlocal operator contains
 information about a whole series of matrix elements of 
local operators of increasing
dimension (but fixed twist), which are encoded in moments of the 
distribution amplitudes. For example:
\begin{equation}
\langle \rho^+(p,\lambda) |\bar u(0)\sigma_{\mu\nu}(D\cdot x)^n d(0)
|0\rangle  = 
-i(e^{(\lambda)}_\mu p_\nu -e^{(\lambda)}_\nu p_\mu)
f_\rho^\perp (ipx)^n \int_0^1 du\, u^n \phi_\perp(u,\mu).
\end{equation}
A renormalization group analysis 
 \cite{exclusive} shows that
for large $n$, moments of the above defined
distribution amplitudes $\phi_{\perp,\parallel}$ behave as\footnote{
Note that a purely perturbative analysis is sufficient to obtain the 
leading
behaviour in $n$, whereas the coefficient of proportionality 
can only be obtained by using nonperturbative methods, see 
\cite{CZreport}.}
\begin{equation}\label{eq:pertbeh}
\int_0^1 du\, u^n \phi(u,\mu) \sim {\rm const}/n^2,
\end{equation}
which corresponds to the following end-point behavior of the 
amplitude for $u\to1$: 
\begin{equation}
  \phi(u,\mu) \sim {\rm const}\cdot(1-u).
\label{eq:end-point}
\end{equation} 
We would like to stress that it is in this place -- using the
information about large $n$ behaviour 
of local operator contributions, related to the end-point behaviour of 
the LC distributions -- that the LC sum rules go 
beyond the traditional SVZ approach. We will discuss this point in
detail in the next section.

The rest of the LC sum rule procedure follows the standard
rules sketched in the last subsection: to suppress contributions of 
higher-twist
$\rho$ meson distributions and to enhance the sensitivity to the
ground $B$ meson state, one performs a Borel transformation in $p_B^2$,
and the final expressions for the sum rules for $B\to\rho$ decay form 
factors
are similar to the 3pt sum rules in (\ref{eq:SR3pt}) apart
from the different expressions on the right-hand side. One thus
obtains (to leading-twist accuracy):
\begin{eqnarray}
\lefteqn{A_1^{B\to\rho}(t)\ =\ \frac{m_b}{f_B(m_B+m_\rho)m_B^2}\,
\exp\left\{
\frac{m_B^2-m_b^2}{M_B^2}\right\} \int_0^1
\frac{du}{u}\,\exp\left\{\frac{\bar
u}{uM_B^2}\,(t-m_b^2-um_\rho^2)\right\}}&&\nonumber\\
& & \Theta[c(u,s_0^B)]\left\{
f_\rho^\perp(\mu)\phi_\perp(u,\mu)\,
\frac{1}{2u}\,(m_b^2-t+u^2m_\rho^2) + f_\rho m_b m_\rho
g_\perp^{(v)}(u,\mu)\right\}\!,\label{eq:LCA1}\\
\lefteqn{A_2^{B\to\rho}(t)\ =\ \frac{m_b(m_B+m_\rho)}{f_Bm_B^2}\,\exp
\left\{
\frac{m_B^2-m_b^2}{M_B^2}\right\} \int_0^1
\frac{du}{u}\,\exp\left\{\frac{\bar
u}{uM_B^2}\,(t-m_b^2-um_\rho^2)\right\}}&&\nonumber\\
& & \left\{ \frac{1}{2}\,f_\rho^\perp(\mu) \phi_\perp(u,\mu)
\Theta[c(u,s_0^B)]
+ f_\rho m_b m_\rho \Phi_\parallel(u,\mu) \left[ \frac{1}{uM_B^2}\,
\Theta[c(u,s_0^B)] + \delta[c(u,s_0^B)]\right]\right\}\!,\label{eq:LCA2}\\
\lefteqn{V^{B\to\rho}(t)\ =\ \frac{m_b(m_B+m_\rho)}{2f_Bm_B^2}\,\exp\left\{
\frac{m_B^2-m_b^2}{M_B^2}\right\} \int_0^1
\frac{du}{u}\,\exp\left\{\frac{\bar
u}{uM_B^2}\,(t-m_b^2-um_\rho^2)\right\}}&&\nonumber\\
& & \left\{ f_\rho^\perp(\mu) \phi_\perp(u,\mu) \Theta[c(u,s_0^B)]
+ \frac{1}{2}\,f_\rho m_b m_\rho g_\perp^{(a)}(u,\mu)\left[ 
\frac{1}{uM_B^2}\,
\Theta[c(u,s_0^B)] + \delta[c(u,s_0^B)]\right]\right\}\!.\makebox[1cm]{\ }
\label{eq:LCV}\end{eqnarray}
with $c(u,s_0^B) = us_0^B-m_b^2+t\bar u-u\bar u m_\rho^2$.

LC sum rules for $A_1$ and $V$ were already obtained in
\cite{ABS}; they slightly differ from the ones given above by the
``surface" terms  $\delta[c(u,s_B^B)]$, which are related to
subtleties in the
continuum subtraction as discussed in App.~B. The LC sum rule
for $A_2$ is new.

\subsection{The Conflict}
\begin{figure}
\centerline{\epsffile{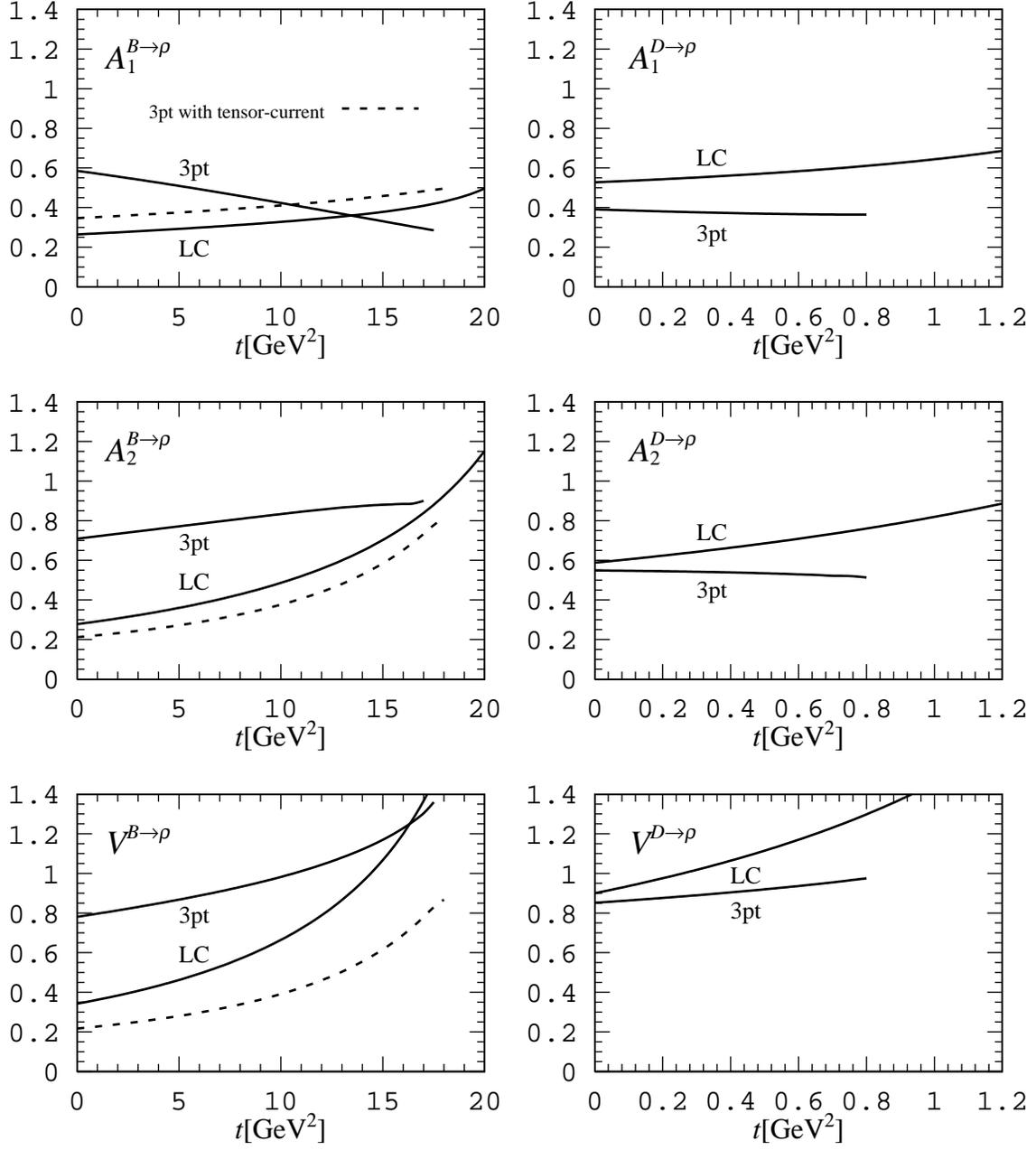}}
\caption[]{Semileptonic form factors of the decays $D\to\rho$ and
$B\to\rho$ from LC sum rules, 
Eqs.~\protect{(\ref{eq:LCA1})--(\ref{eq:LCV})}, and 3pt sum rules,
Eqs.~(\protect{\ref{eq:3ptA1}})--(\protect{\ref{eq:SR3pt}}), evaluated 
with the same input parameters. The
dotted curves illustrate the effect of introducing a different
interpolating current for the $\rho$ meson in the 3pt correlation
function, see Sec.~3.3.}
\end{figure} 
The two approaches described above are rather different and their
comparison should shed light on the actual accuracy of the sum rule
method.  The numerical comparison requires the use of a ``coherent''
set of parameters, so that differences are not introduced (or masked)
by using different inputs. We shall specify our set in detail below;
for the purpose of illustration the particular values are unimportant.
The results for all $B$ and $D$ meson decay form factors from both 3pt
and LC sum rules are shown in Fig.~2.  We see that the results are in
reasonable agreement at large $t$ while there is a disturbing
discrepancy up to a factor two at large recoil \cite{PRD48,ABS}.  The
$t$ dependence also turns out to be very different
\cite{PRD48,ABS}.
 
 Provided no particular advantage or flaw of one method can be found,
this spread of values would necessarily have to be considered as
indicating poor theoretical accuracy of the predictions in this
region.  The further discussion will clarify the reason for this
discrepancy and give strong evidence in favour of the LC sum rule
calculation.  Reasons for the better agreement at small recoil (large
$t$) will also become clear.


\section{Anatomy of the Discrepancy}
\setcounter{equation}{0} An inspection shows that the disagreement
between LC and 3pt sum rules is mainly due to the contribution of the
quark condensate, which dominates the 3pt sum rules at small $t$, cf.\
\cite{PRD48,narrdominance}. To clarify the reason, we give in this
section a detailed calculation of this contribution, and also of the
contribution of the mixed condensate to the 3pt sum rule for the axial
form factor $A_1$.  The result is well known \cite{BBD,PRD48} and the
new point we wish to make here is to rederive it using the sequence of
steps adopted by the LC sum rule approach. This will reveal how the
$\rho$ meson distribution amplitudes are implicitly described in the
3pt approach and also give examples of higher-twist contributions.

\subsection{Three-Point Sum Rule from the Light-Cone Point of View}

We start from the correlation function (\ref{eq:corrga}) and as first
step substitute the heavy quark propagator by its leading-order
perturbative expression:
\begin{eqnarray}
\Gamma^{\mu\nu} & = & i^2\!\!\int d^4x\, d^4y\, e^{-ip_B x +i p y}\,
\int\frac{d^4k}{(2\pi)^4 i} e^{ikx}\frac{1}{m_b^2-k^2}\nonumber\\
&&{}\times \langle 0 | T\Big\{ j_\rho^\nu(y) \bar
u(0)\gamma^\mu(1-\gamma_5) (m_b+\not\!k)i\gamma_5d(x)\Big\} | 0
\rangle.
\label{eq:sample1}
\end{eqnarray}
The product of $\gamma$-matrices in (\ref{eq:sample1}) contains
several terms, corresponding to different invariant structures in
(\ref{eq:corrga}) and to contributions of dimension odd (even)
operators to the OPE.  We choose to consider the axial form factor
$A_1$, and contributions of operators of odd dimension only. To this
end we need to calculate the correlation functions
\begin{eqnarray}
 T^{\nu\mu\alpha} &=& i\int d^4y\, e^{i p y}\, \langle 0 | T\Big\{
  j_\rho^\nu(y) \bar u(0)\sigma^{\mu\alpha}d(x)\Big\} | 0 \rangle,
\label{eq:sample2}
\\ T^{\nu} &=& i\int d^4y\, e^{i p y}\, \langle 0 | T\Big\{
 j_\rho^\nu(y) \bar u(0)d(x)\Big\} | 0 \rangle,
\label{eq:sample2a}
\end{eqnarray}
using the OPE (we recall that $p^2$ is assumed to be Euclidian and
sufficiently large).

Throughout the calculation we imply using Fock-Schwinger gauge.  In a
general (covariant) gauge the heavy quark propagator in external gluon
fields contains the link factor (\ref{eq:gauge-factor}), which has to be
inserted in 
the nonlocal operators in (\ref{eq:sample2}) and (\ref{eq:sample2a})
to make them gauge-invariant, see Sec.~2.2.  In the Fock-Schwinger gauge,
further terms in the expansion of the $b$ quark propagator in
background gluon fields only yield corrections $\sim 1/m_b$ to the sum
rules and for simplicity will not be considered here.  They can easily
be added.

The OPE of the correlation functions (\ref{eq:sample2}) and
(\ref{eq:sample2a}) is straightforward and yields:\footnote{The
perturbative contribution to (\ref{eq:sample2}) and
(\ref{eq:sample2a}) is of order $m_{u,d}$ and will be neglected here.}
\begin{eqnarray}
T^{\nu\mu\alpha} &=& i\langle \bar q q\rangle(p^\alpha
g^{\mu\nu}-p^\mu g^{\alpha\nu})
\Bigg\{(1+e^{ipx})\Big[\frac{1}{p^2}+\frac{1}{3}\frac{m_0^2}{p^4}+
\frac{1}{16} m_0^2x^2\frac{1}{p^2}\Big]
-ipx(1-e^{ipx})\frac{1}{3}\frac{m_0^2}{p^4}\Bigg\} \nonumber\\ &&{} +
i\langle \bar q q\rangle \frac{1}{px} (x^\alpha g^{\mu\nu}-x^\mu
g^{\alpha\nu})(ipx)(1-e^{ipx})\frac{1}{6} \frac{m_0^2}{p^2}
\nonumber\\ &&{}+ i\langle \bar q q\rangle
\frac{1}{px}\frac{p^\nu}{p^2} (p^\alpha x^\mu - p^\mu
x^\alpha)(ipx)(1-e^{ipx})\frac{1}{6} \frac{m_0^2}{p^2}\,,
\label{eq:sample3}\\
T^\nu &=& i\langle \bar q q\rangle \left(p^\nu(px)-x^\nu
p^2\right)(1+e^{ipx})\frac{1}{6} \frac{m_0^2}{p^4} -\langle \bar q
q\rangle\,
\frac{p^\nu}{p^2}(1-e^{ipx})\Big(1+\frac{1}{16}m_0^2x^2\Big).
\label{eq:sample3a}
\end{eqnarray}
Here $m_0^2= \langle \bar q g\sigma G q\rangle/\langle \bar q
q\rangle$.  Note that $p_\nu T^{\nu\mu\alpha}=0$, while $T^\nu$
contains a contact term.  Substituting (\ref{eq:sample3}) and
(\ref{eq:sample3a}) into (\ref{eq:sample1}), taking the remaining
integrals and performing Borel transformations in $p_B^2$ and $p^2$,
respectively, we reproduce the contributions of quark and mixed
condensate to the 3pt sum rule for $A_1$ in Ref.~\cite{PRD48}, except
for the neglected contribution of the diagram with the gluon emitted
from the $b$ quark line:
\begin{eqnarray}
A_1^{B\to\rho}(t) & = & \frac{
m_b}{f_Bf_\rho(m_B+m_\rho)m_B^2m_\rho}\,\exp\left\{
\frac{m_B^2-m_b^2}{M_B^2}+\frac{m_\rho^2}{M_\rho^2}\right\}
\left\{-\langle\bar q q\rangle \frac{1}{2}(m_b^2-t)\right.
\nonumber\\ &+&\left.\langle \bar q g\sigma G q\rangle\left[
-\frac{7}{24}+\frac{m_b^2(m_b^2-t)}{8M_B^4}-\frac{m_b^2-t}{6M_\rho^2}
-\frac{3m_b^2-5t}{24M_B^2} +
\frac{(m_b^2-t)^2}{6M_B^2M_\rho^2}\right]\right\}.
\label{eq:patricia}
\end{eqnarray}
We emphasize that the derivation sketched above is entirely within the
traditional QCD sum rule approach, although the sequence of steps may
seem unusual. A related discussion for the
$\pi\gamma\gamma^*$ transition form factor can be found in \cite{RadRus}.

We now rewrite this answer in terms of contributions of $\rho$ meson
distribution amplitudes.  To this end, we separate the $\rho$ meson
contribution to $ T^{\nu\mu\alpha}(p)$,
\begin{equation}
 T^{\nu\mu\alpha}(p) = \langle 0 |j_\rho^\nu| \rho^+(p,\lambda)\rangle
\frac{1}{m_\rho^2-p^2} \langle \rho^+(p,\lambda)|\bar
u(0)\sigma^{\mu\alpha}d(x)|0\rangle +\ldots,
\label{eq:sample4}  
\end{equation}    
and, similarly, the one to $T^\nu$.  The first matrix element is
proportional to the decay constant $f_\rho$, while the second one, by
definition, gives $\rho$ meson distribution amplitudes in the fraction
of momentum carried by the quark.  An inspection of (\ref{eq:sample3})
and (\ref{eq:sample3a}) suggests to introduce the following
distributions:
\begin{eqnarray}
\langle \rho^+(p,\lambda)|\bar u(0)\sigma_{\mu\alpha}d(x)|0\rangle &=&
-if_\rho^\perp(e^{*(\lambda)}_\mu p_\alpha -e^{*(\lambda)}_\alpha
p_\mu) \int_0^1 du\, e^{iupx} [\phi_\perp(u) +x^2 \psi^{(1)}(u)]
\nonumber\\ &&{}+f_\rho^\perp
(e^{*(\lambda)}_\mu x_\alpha -e^{*(\lambda)}_\alpha x_\mu) \int_0^1
du\, e^{iupx} \psi^{(2)}(u) \nonumber\\ &&{}+i f_\rho^\perp
(e^{*(\lambda)}\cdot x) (x_\mu p_\alpha -x_\alpha p_\mu) \int_0^1 du\,
e^{iupx} \psi^{(3)}(u),
\label{eq:sample6}
\\ \langle \rho^+(p,\lambda)|\bar u(0)d(x)|0\rangle &=&
-if_\rho^\perp(e^{*(\lambda)}\cdot x) \int_0^1 du\, e^{iupx}
\psi^{(4)}(u).
\label{eq:sample6a}
\end{eqnarray}
After a Borel transformation of (\ref{eq:sample3}) and
(\ref{eq:sample4}) in $p^2$ we get the explicit expressions
\begin{eqnarray}
 \phi_\perp(u) &=&\frac{-\langle\bar q q\rangle}{m_\rho f_\rho
f_\rho^\perp} e^{m^2_\rho/M_\rho^2}\Bigg\{
\left(1-\frac{1}{3}\frac{m_0^2}{M_\rho^2}\right)[\delta(u)+\delta(1-u)]
-\frac{1}{3}\frac{m_0^2}{M_\rho^2}\frac{d}{du}[\delta(u)-\delta(1-u)]
\Bigg\},
\nonumber\\[-15pt]\label{eq:phidelta}\\ \psi^{(1)}(u) &=&
\frac{-\langle\bar q q\rangle}{m_\rho f_\rho f_\rho^\perp}
e^{m^2_\rho/M_\rho^2} \frac{1}{16}\,m_0^2[\delta(u)+\delta(1-u)], \\
\psi^{(2)}(u) &=&\frac{-\langle\bar q q\rangle}{m_\rho f_\rho
f_\rho^\perp} e^{m^2_\rho/M_\rho^2}
\frac{1}{6}\,m_0^2[\delta(u)-\delta(1-u)],\\ \psi^{(3)}(u) &=& 0\,,
\label{eq:sample7}
\end{eqnarray}
where $M_\rho^2\approx (1-2)\,$GeV$^2$ is the Borel parameter.  Note
that the expansion goes in derivatives of the
$\delta$-function. A similar expansion for the twist 2
distribution amplitude was obtained in \cite{radmikh}.

Similarly, from the expansion (\ref{eq:sample3a}) we deduce
\begin{equation}
  \psi^{(4)}(u) = \frac{\langle\bar q q\rangle}{m_\rho f_\rho
 f_\rho^\perp} e^{m^2_\rho/M_\rho^2} \frac{1}{6}\,m_0^2
 [\delta(u)+\delta(1-u)].
\label{eq:sample8}
\end{equation}  

Substituting (\ref{eq:sample6}) and (\ref{eq:sample6a}) in
(\ref{eq:sample4}) and (\ref{eq:sample1}), taking the integrals and
performing a Borel transformation in $p_B^2$, we get a typical LC sum
rule:
\begin{eqnarray}
\lefteqn{A_1^{B\to\rho}(t)\ =\
\frac{m_bf_\rho^\perp}{f_B(m_B+m_\rho)m_B^2}\,
\exp\left\{\frac{m_B^2-m_b^2}{M_B^2}\right\} \int_0^1
\frac{du}{u}\,\exp\left\{\frac{\bar
u}{uM_B^2}\,(t-m_b^2-um_\rho^2)\right\}}&&\nonumber\\ & &
\times\Bigg\{\frac{1}{2u}\,(m_b^2-t+u^2m_\rho^2) \Bigg[
\phi_\perp(u)-\frac{4}{u M_B^2}\left(1 +\frac{m_b^2}{u M_B^2}\right)
\psi^{(1)}(u) +\frac{2}{u M_B^2}\psi^{(3)}(u) \Bigg] \nonumber\\ & &
{}+\frac{2}{u}\left[1+\frac{2m_b^2}{u M_B^2}\right] \psi^{(1)}(u)
+\left[1+\frac{2m_b^2}{u M_B^2}\right]\psi^{(2)}(u)
-\frac{1}{u}\psi^{(3)}(u)-\psi^{(4)}(u)\Bigg\},
\label{eq:sample9}
\end{eqnarray}
where we have changed variables $u\rightarrow 1-u$ to be consistent
with (\ref{eq:LCA1}).  To save space we have not shown the continuum
subtraction.  Note that the leading-twist contribution of the
distribution amplitude $\phi_\perp$ coincides with the corresponding
contribution in (\ref{eq:LCA1}); the extra terms $\psi^{(i)}$ are
higher-twist corrections, not taken into account in
(\ref{eq:LCA1}).\footnote{ The contribution $\sim g_\perp$ in
(\ref{eq:LCA1}) would correspond to terms in (\ref{eq:sample1}) with
an odd number of $\gamma$-matrices, which we have not considered
here.}

On the other hand, further substituting in (\ref{eq:sample9}) the
above expressions for the distribution amplitudes and suppressing
terms $\sim m_\rho^2$ unless they get divided by $M_\rho^2$, we come
back to (\ref{eq:patricia}). The quark condensate contribution in
(\ref{eq:patricia}) appears as a contribution of the leading-twist
distribution $\phi_\perp$, while the mixed condensate terms contain
contributions from both leading- and higher-twist. In particular, for
the expression in square brackets in (\ref{eq:patricia}) we find the
following decomposition:
\begin{eqnarray}
\left[\frac{(m_b^2-t)^2}{6M_B^2M_\rho^2} -\frac{m_b^2-t}{6M_\rho^2}
+\frac{m_\rho^2}{M_\rho^2}
\left(\frac{1}{6}+\frac{m_b^2-t}{3M_B^2}\right)\right]_{\phi_\perp}\!\!\!
& + & \left[\frac{m_b^2(m_b^2-t)}{8M_B^4}+\frac{m_b^2-t}{8 M_B^2}-
\frac{m_b^2}{4M_B^2} -\frac{1}{8}\right]_{\psi^{(1)}} \nonumber\\
{}+\left[\frac{1}{6}+\frac{m_b^2}{3M_B^2}\right]_{\psi^{(2)}}\!\!\!  &
- & \left[ \vphantom{\frac{m_b^2}{3M_B^2}}
\frac{1}{6}\right]_{\psi^{(4)}},
\label{eq:decomposition}
\end{eqnarray}
where $[\ldots]_{\phi_\perp}$ indicates that this term originates from
the distribution $\phi_\perp$, etc.  As it stands, this expression
does not yet agree with (\ref{eq:patricia}), the reason being that the
Borel transformation in the $\rho$ meson channel was applied in a
slightly different way.  It is possible to show that in order to
reproduce the 3pt sum rule one has to substitute $m_\rho^2\to
-M_\rho^2$, after which the expressions indeed coincide
literally\footnote{ There is a subtlety in treating the terms
proportional to $p_\alpha/p^4$ in the first line in
(\ref{eq:sample3}): $p_\alpha$ gets contracted with $p_B$ and yields a
factor $p_B^2+p^2-t$. Using the dispersion relation first in the
$\rho$ meson channel like in (\ref{eq:sample4}) then implies that
$p^2$ is substituted by $m_\rho^2$, while in the standard procedure it
gives $ - M_\rho^2$. Ambiguities of this type are intrinsic for the
sum rule method.}.

\subsection{Short Distance Expansion and Light-Cone Distribution 
Amplitudes}  
\begin{figure}
\centerline{\epsffile{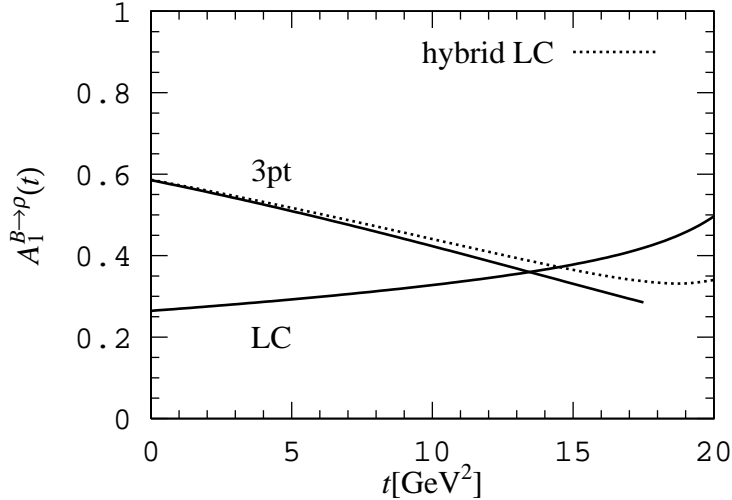}}
\caption[]{$A_1$ from the LC sum rule (\protect{\ref{eq:LCA1}}) (solid
line LC), the 3pt sum rule (\protect{\ref{eq:3ptA1}}) (solid line
3pt), and a ``hybrid LC'' sum rule, in which the leading-twist
distribution amplitude $\phi_\perp$ is replaced by the naive expansion
in $\delta$-functions, (\protect{\ref{eq:phidelta}}).}
\end{figure}

The new input made by the LC sum rules is to argue that the
$\delta$-function type shape of LC distributions, concentrated at
$u=0$ and $u=1$, is {\em qualitatively} wrong. In particular, instead
of the expression in (\ref{eq:phidelta}), it is suggested to use the
distribution amplitude
\begin{equation}
\phi_\perp(u,\mu) = 6u(1-u) \left[ 1+
a_2^\perp(\mu)C^{3/2}_2(2u-1)\right].
\label{eq:rhoWF}
\end{equation}
Here $C_2^{3/2}(x) = (15x^2-3)/2$ is the second order Gegenbauer
 polynomial; the coefficient $a^\perp_2$ was estimated to be $0.2\pm
 0.1$ \cite{rhoWF}.  Eq.~(\ref{eq:rhoWF}) is clearly very different
 from (\ref{eq:phidelta}).  Where does it come from and what is wrong
 with (\ref{eq:phidelta})?

The distributions (\ref{eq:phidelta})--(\ref{eq:sample8}) are just the
QCD sum rules for the correlation functions (\ref{eq:sample2}) and
(\ref{eq:sample2a}).  Their deficiency becomes apparent when they are
rewritten in terms of moments. For the leading-twist distribution we
find (cf. \cite{rhoWF}):
\begin{equation}
  \int_0^1 du\,(2u-1)^n \phi_\perp(u) = \frac{-\langle\bar q
  q\rangle}{m_\rho f_\rho f_\rho^\perp}\,e^{m_\rho^2/M_\rho^2}
  \left(1+ (-1)^n\right)
  \left(1-\frac{m_0^2}{3M_\rho^2}\,(2n+1)\right).
\label{eq:rhoWFmom}
\end{equation}
Note that the contribution of the mixed condensate is enhanced by a
factor $n$. This enhancement is of generic nature: contributions of
higher dimension to the OPE will be acompanied by increasing powers of
$n$ so that the sum rule blows up for large moments and cannot be
used. This signals the break-down of local OPE for higher moments of
distribution amplitudes.  Extensive studies \cite{CZreport} have
demonstrated that QCD sum rules of type
(\ref{eq:phidelta})--(\ref{eq:sample8}) can be applied to estimate the
two first moments only, $n=0$ and $n=2$, i.e.\ the normalization and
width of the distribution amplitudes, but fail to describe higher
moments, i.e.\ the shape of the distribution close to the end-points.
Information on the shape can, however, be obtained from another
source, namely the behaviour of distribution amplitudes under the
renormalization group \cite{exclusive}.  The major result is that
$\phi_\perp$ approaches $6u(1-u)$ at large virtualities and that the
corrections can be systematically expanded in Gegenbauer polynomials
$C^{3/2}_n(2u-1)$. Combining this expansion with estimates of the
first two moments by QCD sum rules one obtains the expression
(\ref{eq:rhoWF}).

In fact, the particular sum rule in (\ref{eq:rhoWFmom}) is not
accurate enough even for $n=0,2$, and in practice one uses different
sum rules, see Ref.~\cite{rhoWF} for a detailed discussion.

To illustrate that the shape of the leading-twist distribution is
indeed of crucial importance, we have plotted in Fig.~3 the form
factor $A_1^{B\to \rho}(t)$, calculated in several different ways.
The solid curve, labelled LC, shows the LC sum rule (\ref{eq:LCA1})
with realistic distribution amplitudes. The dotted line is obtained
using the same sum rule (\ref{eq:LCA1}), but with the distribution
amplitude $\phi_\perp$ replaced by the expression (\ref{eq:phidelta});
it is very close to the solid line showing the 3pt sum rule result.
The ``dominance'' of the quark condensate \cite{narrdominance} in the
3pt sum rule thus happens to be an artifact of the short-distance
expansion extrapolated beyond the region of its validity.

The ideal agreement of the dotted curve in Fig.~3 with the 3pt sum
rule result at $t=0$ is in fact coincidental and is due to a mutual
cancellation of two effects.  First, in addition to contributions of
operators of odd dimension, the 3pt sum rule contains a perturbative
term, a contribution of four-quark operators of dimension 6, and of
the gluon condensate.  These contributions correspond to the terms
with an odd number of $\gamma$-matrices in (\ref{eq:sample1}), which we
have not considered, and have their counterpart in the LC sum rule in
the contribution of the distribution $g_\perp$ (up to higher-twist
terms).  The difference between the two approaches is small in this
case, the reason being that repeating the above procedure one would
deal with the correlation function of $j_\rho^\nu$ with a nonlocal
vector current. In contrast to (\ref{eq:sample2}),
(\ref{eq:sample2a}) it has a large perturbative contribution and
the OPE goes in condensates of even dimension.
Extracting the distribution amplitude as outlined above would yield a
smooth distribution $\sim u(1-u)$, slightly corrected by
$\delta$-function type contributions of the gluon and four-quark
condensates. These latter contributions are small, so that $g_\perp$
as implicitly used in the 3pt sum rules is not very different from its
``true'' behaviour. Hence, the numerical results are close.

Second, the present version of the LC sum rule neglects contributions
of higher-twist.  To estimate their effect one can apply the methods
of Ref.~\cite{BF2} to determine the shape of the distributions
$\psi^{(k)}(u)$ at large scales, i.e.\ their asymptotic form, and use
the sum rules (\ref{eq:sample7}) to estimate the normalization. We get
\begin{eqnarray}
  \psi^{(1)}(u) &=& \kappa_1 \cdot 30 u^2(1-u)^2, \nonumber\\
\psi^{(2)}(u) &=& \kappa_2 \cdot 30 u(1-u)(1-2u), \nonumber\\
\psi^{(4)}(u) &=& \kappa_4 \cdot 6 u(1-u)\label{eq:modelHT}
\end{eqnarray}
with
\begin{equation}\label{eq:modelkappa}
 \kappa_1 = m_0^2/16,\quad\kappa_2 = m_0^2/6,\quad \kappa_4 =
-m_0^2/6.
\end{equation}
\begin{figure}
\centerline{\epsffile{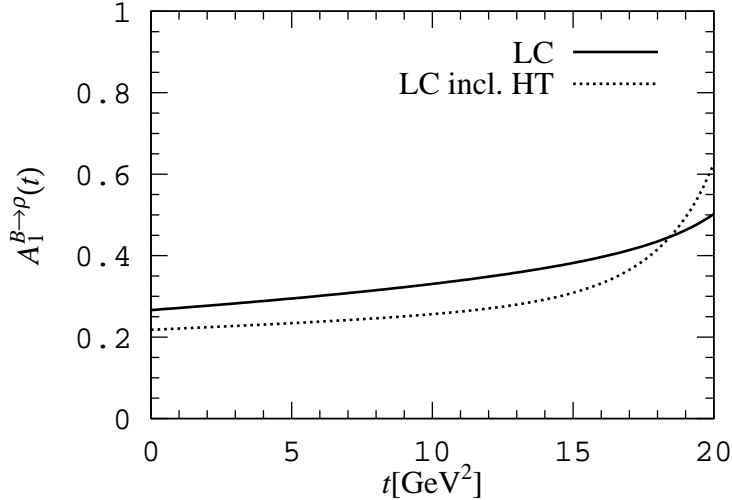}}
\caption[]{$A_1$ from the LC sum rules (\protect{\ref{eq:LCA1}}) and
(\protect{\ref{eq:sample9}}), with the distribution amplitudes
(\protect{\ref{eq:modelHT}}), (\protect{\ref{eq:modelkappa}}).}
\end{figure}
In Fig.~4 we plot $A_1^{B\to\rho}$ from Eq.~(\ref{eq:sample9}) using
these distributions and including continuum subtraction. For
comparison we also show the leading-twist LC sum rule
(\ref{eq:LCA1}). The correction turns out to be negative and lowers
the leading-twist result by about 15\% for $t<15\,{\rm GeV}^2$. These
results are, however, only indicative on the size of higher-twist
corrections, the detailed study of which goes beyond the tasks of this
paper.
 
If the ``naive'' description of distribution amplitudes by the usual
sum rule method is that deficient, the question arises if this
approach still works for form factors of $D$ mesons, as used e.g.\ in
\cite{BBD}.  The formal answer is clear from the structure of LC sum
rules: the distribution amplitudes are integrated with a smooth weight
function over a constrained region of the momentum fraction $u$.  If
the mass of the heavy meson is not very large compared to the typical
hadronic scale 1~GeV, then the integration region is large and only
gross characteristics of the distribution amplitudes matter, i.e.\
their normalization and width.  These are given correctly by the sum
rules, and the approach works well.  If, on the other hand, the mass
of the heavy meson is much larger than 1~GeV, as it is the case with
$B$ mesons, and if the momentum transfer to the leptons is small, then
the integration region shrinks to the narrow interval $1-u\sim
O(1/m_b)$, the precise behaviour of the distribution amplitude at
$u\to 1$ becomes important, and the standard approach fails.

\begin{figure}
\centerline{\epsffile{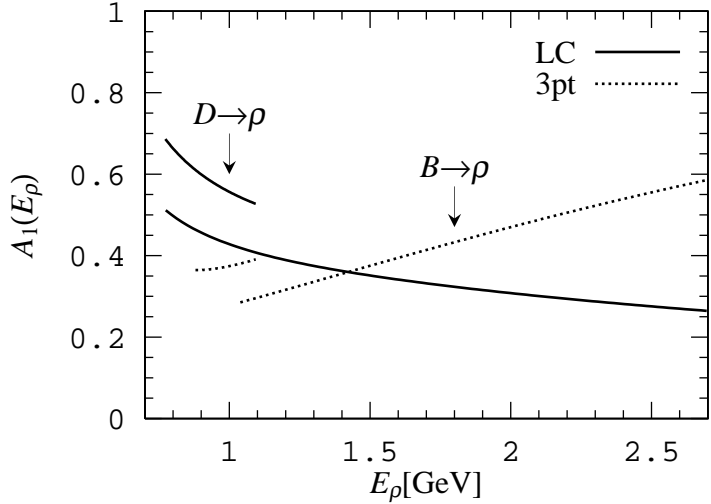}}
\caption[]{$A_1$ as function of the $\rho$ meson energy $E_\rho$ from
LC and 3pt sum rules for both $D\to\rho$ and $B\to\rho$
transitions. For $D\to\rho$ transitions, the physical region $t>0$
corresponds to $\rho$ meson energy up to 1.1~GeV.}
\end{figure}
The physical parameter that matters is, however, not the heavy meson
mass, but the $\rho$ meson energy $E_\rho$ in the decaying $B(D)$ rest
frame: $E_\rho= (m_{B,D}^2+m_\rho^2-t)/(2m_{B,D})$. Zero recoil
corresponds to $E_\rho =m_\rho$; in the physical regime $t>0$,
$E_\rho$ runs up to 2.7~GeV and 1.1~GeV in $B$ and $D$ decays,
respectively. In Fig.~5 we show the form factors $A_1(E_\rho)$ for
both $B$ and $D$ mesons. The behaviour is very similar, and in both
channels 3pt and LC sum rules agree very well for $E_\rho \approx
1.4\,$GeV. For $D$ mesons, this is outside the physical region for the
decay.

\subsection{Possible Remedy: the Tensor-Current?}
To conclude this section, we would like to demonstrate that the
 ``dominance'' of the quark condensate is no intrinsic feature of 3pt
 sum rules. To this end we recall that one has some freedom in the
 choice of the interpolating field for the considered particles:
 although for the $\rho$ meson the vector current is the most
 convenient one, it is by no means the only one. In particular one can
 choose the tensor current $j_\rho^{\nu\sigma} = \bar d
 \sigma^{\nu\sigma} u$ instead and calculate the $B\to\rho$ form
 factors from the correlation function
\begin{equation}
\Sigma^{\mu\nu\sigma} = i^2\int d^4x d^4y e^{-ip_Bx+ip_\rho
y}\,\langle0|Tj^{\nu\sigma}_\rho(y)
(V-A)^\mu(0)j_B^\dagger(x)|0\rangle\label{eq:sigmacorr}.
\end{equation}
In App.~B we give the corresponding OPE including terms up to
dimension six.  Due to the particular $\gamma$-matrix structure, the
contribution of the quark condensate to $A_2$ and $V$ vanishes and is
small for $A_1$. We have displayed the corresponding form factors
already in Fig.~2.  They differ distinctively from the results of the
original 3pt sum rules and are much closer numerically to the LC sum
rules. Nevertheless it would be inappropriate to conclude that the
above correlation function is ``better'' than (\ref{eq:corrga}): it
suffers from exactly the same problem as the original correlation
function to describe correctly the shape of the $\rho$ meson
distribution amplitudes near the end-points in $u$.  It is only that
this failure is less ``visible'' for the given values of the $b$ quark
mass and the considered range in $t$.  The problem is now shifted to
the contribution of the mixed condensate, which starts to domininate
$A_1(0)$ at large $m_b$ and eventually overgrows all other terms.
Numerically, however, the effect is much less significant at
$m_b\sim$~5~GeV.  This improvement comes at the price that the tensor
current couples also to positive parity $1^{+-}$ states which contaminate
the contribution of the $\rho$ meson, so that the accuracy of these
sum rules is not very high.  Another possibility to achieve a
similarly ``favourable'' rearrangement of power corrections would be
to use the axial-vector instead of the pseudo-scalar current for the
$B$ meson.


\section{The Heavy Quark Limit}
\setcounter{equation}{0}

The behavior of form factors in the limit $m_b\to\infty$ is of
considerable theoretical and practical interest.  Taking the heavy
quark limit in the sum rules is straightforward, by rescaling the sum
rule parameters in the following way (see e.g.\ \cite{BBBD}):
\begin{equation}
M^2\to 2 m_b\tau, \qquad s_0^B\to m_b^2 + 2 m_b \omega_0,
\end{equation}
where $\tau$ and $\omega_0$ are of order 1~GeV.
 
One should distinguish between different regions of momentum 
transfer. First, consider $m_b^2-t\sim O(m_b)$, i.e.\ small recoil,
energy of the outgoing $\rho$ meson of order 1~GeV.  Then both the 3pt
and the LC sum rules fulfil the scaling laws predicted by heavy quark
effective theory \cite{IW}:
\begin{equation}
A_1(t\approx t_{max}) \sim 1/\sqrt{m_b}, \quad A_2(t\approx t_{max})
\sim \sqrt{m_b}, \quad V(t\approx t_{max}) \sim \sqrt{m_b}.
\label{eq:scaling}
\end{equation}
In this regime, the integration over the quark momentum fraction in LC
sum rules comprises the region $1-u\sim O(1)$, so that only width and
normalization of the distributions are important. Hence 3pt and LC sum
rules are expected and indeed give comparable results, see Figs.~2,3.

More interesting, however, is the behaviour near maximum recoil,
$t\approx 0$. Here we find that in the 3pt sum rules approach the
limit $m_b\to\infty$ cannot be taken since higher order terms in the
OPE are accompanied by increasing powers of $m_b$.  {}From the
``light-cone point of view'' this inconsistency arises because at
large recoil the soft contributions to the form factors pick up a tiny
region of momentum fraction $1-u \sim O(1/m_b)$ and thus the details
of the shape of the $\rho$ meson distribution amplitudes, wrongly
described by 3pt sum rules, enter decisively.

On the contrary, LC sum rules at $t=0$ have a well-defined heavy quark
limit \cite{ABS} and scale as $1/m_b^{3/2}$.  Explicitly, making the
change of variables $\omega = (1-u)m_b/2$, one finds (with
$\hat f = f_B\sqrt{m_b}$ and $\bar\Lambda= m_B-m_b$):
\begin{eqnarray}
\hat f A_1(0) & = & -\frac{2}{m_b^{3/2}}\,e^{\bar\Lambda/\tau}
\int_0^{\omega_0} d\omega\,e^{-\omega/\tau} \left[f_\rho^\perp
\omega\phi'_\perp(1) - f_\rho m_\rho g_\perp^{(v)}(1)
\right],\nonumber\\ 
\hat f A_2(0) & = &
-\frac{2}{m_b^{3/2}}\,e^{\bar\Lambda/\tau} \int_0^{\omega_0}
d\omega\,e^{-\omega/\tau} \left[f_\rho^\perp
\omega\phi'_\perp(1) + f_\rho m_\rho \Phi'_\parallel(1)
\right],\nonumber\\ 
\hat f V(0) & = &
-\frac{2}{m_b^{3/2}}\,e^{\bar\Lambda/\tau} \int_0^{\omega_0}
d\omega\,e^{-\omega/\tau} \left[f_\rho^\perp
\omega\phi'_\perp(1) + \frac{1}{4}\,f_\rho m_\rho g'^{(a)}_\perp(1)
\right].
\end{eqnarray}
{}From the relations (\ref{eq:bigphi}) and (\ref{WW1}) it follows that
to our accuracy
\begin{equation}
A_1(0) \equiv A_2(0) \equiv V(0)
\label{eq:equality}
\end{equation}
in the heavy quark limit. This agrees with the findings of Ref.~\cite{stech}.
It is instructive to check that the above scaling relations are not
spoiled by higher-twist corrections.  The twist 4 part of
(\ref{eq:sample9}) becomes in the heavy quark limit:
\begin{eqnarray}
\hat{f}A_1^{\rm twist 4}(0) =
\frac{1}{m_b^{5/2}}\,e^{\bar\Lambda/\tau} \int_0^{\omega_0}
d\omega\,e^{-\omega/\tau}\,\left\{ -\frac{4\omega^2}{\tau^2}\,
\psi''^{(1)}(1) - \frac{4\omega}{\tau} \, \psi'^{(2)}(1)
-\frac{\omega}{\tau}\,\psi'^{(3)}(1) \right\},
\end{eqnarray}
where $\psi'(u) =(d/du)\psi(u)$ and we used that all $\psi$-functions
vanish at $u=1$.  It is seen that higher-twist corrections are in fact
down by an extra power of $m_b$, cf.\ the discussion of the pion form
factor in the third of Refs.~\cite{softpion}.

We recall that the heavy quark mass dependence of form factors at zero
recoil is of vivid interest for lattice calculations. Due to
restrictions on computer power and performance, reliably simulable
quark masses are of order $\sim 2\,$GeV and the results have to be
extrapolated to the physical beauty quark mass.  In this respect, we
would like to add a word of caution about using the asymptotic scaling
law $1/m^{3/2}$ since this limit is only approached very slowly \cite{ABS}.
To get a ball-park estimate of the next-to-leading order corrections
we calculated the form factors using LC sum rules varying the $b$ quark
mass in the limits \mbox{(1--10)~GeV} and using the 
scaling (\ref{eq:scaling}) of
the sum rule parameters.  We then fit by a quadratic
polynomial in the inverse {\em meson}\/ mass $m_B=m_b+\bar\Lambda$,
$\bar\Lambda =500$~MeV \cite{BBBD}. The results are (we show the
leading $1/m_B$ corrections only):
\begin{eqnarray}
m_B^{3/2}A_1(0) &=& 5.6\,{\rm GeV}^{3/2}
\left(1-\frac{2.4\, {\rm GeV}}{m_B}+\ldots\right),\nonumber\\ 
m_B^{3/2}A_2(0) &=& 5.6\,{\rm GeV}^{3/2}\left(1-\frac{2.1\, {\rm
GeV}}{m_B}+\ldots\right), \nonumber\\ 
m_B^{3/2}V(0) &=& 5.9\,{\rm GeV}^{3/2}
\left(1-\frac{1.5\, {\rm GeV}}{m_B}+\ldots\right).
\end{eqnarray}
The constants in front of the brackets are almost equal, as expected 
from (\ref{eq:equality}). Note the large terms in $1/m_B$.
 
Finally, one can consider the region of very small recoil $m_b^2-t\sim
O(1\,$GeV).  This region is generally difficult for QCD sum rule
treatment since one gets more sensitive to contributions of large
distances in the ``$t$-channel''. An inspection of
(\ref{eq:decomposition}) shows that in this limit the leading-twist
contributions of dimension~5 are smaller than those of higher-twist,
which may be considered as an indication that 3pt sum rules become
more reliable than LC sum rules at very large $t$.


\section{Numerical Analysis}
\setcounter{equation}{0}

\noindent We now turn to the numerical evaluation of the LC sum rules
(\ref{eq:LCA1})--(\ref{eq:LCV}). Let us first define the relevant
observables.

\subsection{Kinematics}

With the standard decomposition for the $B\to\rho$ transition matrix
element (\ref{eq:ME}) the spectrum with respect to the electron energy
$E$ reads:
\begin{eqnarray}
\lefteqn{\frac{d\Gamma(\bar{B}^0\to\rho^+ e^-\bar \nu)}{dE} = }
\nonumber\\ &
= & \frac{G_F^2|V_{ub}|^2}{128\pi^3m_B^2}\int\limits_0^{t_{max}}\!\!
dt \, t \{(1 - \cos \theta )^2 H_-^2 + (1 + \cos \theta)^2 H_+^2 + 2
(1 - \cos^2 \theta ) H_0^2\},\makebox[1.4cm]{}
\end{eqnarray}
with the helicity amplitudes
\begin{eqnarray}
H_\pm & = & (m_B + m_\rho) A_1(t) \mp \frac{\lambda^{1/2}}{m_B+m_\rho}
V(t),\\ H_0 & = & \frac{1}{2m_\rho\sqrt{t}}\left\{ (m_B^2 - m_\rho^2 -
t) (m_B + m_\rho) A_1(t) - \frac{\lambda}{m_B +
m_\rho}\,A_2(t)\right\},
\end{eqnarray}
where the indices denote the polarization of the $\rho$. $\lambda$ is
defined as
\begin{equation}
\lambda = (m_B^2+m_\rho^2-t)^2-4 m_B^2 m_\rho^2.
\end{equation}
$t_{\rm max}$, the maximum value of $t$ at fixed electron energy, is
given by
\begin{equation}
t_{\rm max} = 2E \left( m_B - \frac{m_\rho^2}{m_B-2E} \right).
\end{equation}
$\theta$ is the angle between the $\rho$ and the charged lepton in the
$(e^- \bar\nu)$ CM system and given by
\begin{equation}
\cos \theta = \frac{1}{\lambda^{1/2}} \,( m_B^2 -m_\rho^2 + t - 4 m_B
E).
\end{equation}
The spectrum with respect to $t$ reads:
\begin{equation}
\frac{d\Gamma(\bar{B}^0\to\rho^+ e^-\bar \nu)}{dt} =
\frac{G_F^2|V_{ub}|^2}{192\pi^3m_B^3}\,\lambda^{1/2}\,t\left(H_0^2 +
H_+^2 + H_-^2\right).
\end{equation}
We also introduce the notations $\Gamma_T$ and
$\Gamma_L$ for the partial decay rates where the final state $\rho$
 is transversely or longitudinally polarized.

{}From the specific structure of the helicity amplitudes it follows
that at small $t$ the produced $\rho$ mesons are predominantly
longitudinally polarized; for $t=0$ only longitudinally polarized
$\rho$ are produced. At large $t$, on the other hand, the
contribution of $A_2$ and $V$ to the decay rate is suppressed, since
$\lambda$ has a zero at $t_{\rm max}$.

\subsection{Input Parameters}

The decay constant $f_\rho$ is measured experimentally \cite{PDG}:
\begin{equation}
f_\rho = (205\pm 10)\,{\rm MeV},
\end{equation}
while existing information on $f_\rho^\perp$ comes from QCD sum rules.
In the following we use \cite{rhoWF}
\begin{equation}
f_\rho^\perp(1\,{\rm GeV}) = (160\pm 10)\,{\rm MeV}.
\end{equation}
The $\rho$ meson leading twist distribution amplitudes
$\phi_\parallel$ and $\phi_\perp$ have been recently reexamined in
\cite{rhoWF}.  We use
\begin{equation}\label{eq:phiparperp}
\phi_{\parallel,\perp}(u,\mu) = 6u(1-u) \left[ 1+
a_2^{\parallel,\perp}(\mu)C^{3/2}_2(2u-1)\right]
\end{equation}
with $a_2^\parallel(1\,{\rm GeV}) = 0.18\pm 0.10$
\cite{CZreport,rhoWF} and $a_2^\perp(1\,{\rm GeV}) = 0.2\pm 0.1$
\cite{rhoWF}, as already mentioned in Sec.~3.2.

The value of the $b$ quark (pole) mass $m_b$ is somewhat
controversial, with estimates varying from 4.6 to 5.1 GeV. This large
range, however, probably overestimates the actual uncertainty and
rather reflects that the pole mass has to be nonperturbatively defined
and that suitable definitions (and values) depend on the application.
In this paper we use
\begin{equation}
    m_b = (4.8\pm 0.1)\, {\rm GeV},
\end{equation}
which, we believe, is a fair estimate.

The decay constant $f_B$ was calculated in QCD sum rules and on the
lattice, with a world average of about 180 MeV (see,
e.g.\ \cite{latticefB}).  It was found, however, that within the QCD
sum rule approach $f_B$ receives large radiative corrections, which
increase its value by 30 to 60 MeV \cite{BBBD}. Since similar
radiative corrections have not been calculated for the sum rules for
form factors, we think that it more consistent to use the lower value of
$f_B$ as it is obtained from the sum rules without radiative
corrections, see also \cite{ABS}.  In practice, we simply substitute
$f_B^2$ by the corresponding sum rule with the same values of all
parameters; this has an additional advantage of reducing considerably
the $b$ quark mass dependence.  In fact, there are arguments suggesting
that radiative corrections tend to cancel between $f_B$ and the form
factors.  This cancellation was indeed observed for $B\to D^*$
transitions \cite{IWrad} and for the $B$ meson matrix element of the
kinetic energy operator \cite{BB}.  An explicit calculation of the
radiative corrections to LC sum rules would, however, be very welcome.

For the values of the condensates we use
\begin{eqnarray}
\langle \bar q q \rangle(1\,{\rm GeV}) & = & -(245\pm 10)\,{\rm
MeV}^3,\nonumber\\ \left\langle \frac{\alpha_s}{\pi}\,G^2\right\rangle
& = & (0.012\pm 0.006)\,{\rm GeV}^4,\nonumber\\ \langle\bar q g\sigma
G q \rangle (1\,{\rm GeV}) & = & 0.65\,{\rm GeV}^2\cdot\langle \bar q
q \rangle(1\,{\rm GeV}),\nonumber\\ \langle\alpha_s\bar q q\rangle^2 &
= & 0.56\cdot (-0.245)\,{\rm GeV}^6.
\end{eqnarray}
They enter the 3pt sum rules explicitly, and the LC sum rules
implicitly, via estimates of the parameters of the distribution
amplitudes \cite{rhoWF} and of $f_B$.

We assume values of the continuum thresholds for $\rho$ and $B$ mesons
$s_0^\rho =1.5$~GeV and $s_0^B =35, 34, 33 $ GeV$^2$ for $m_b=4.7,
4.8, 4.9 $ GeV, respectively.  The working region of Borel parameters
in 3pt sum rules is taken to be $M^2_\rho\approx\,$(1--2)~GeV$^2$ for
$\rho$ mesons and $M^2_B\approx\,$(5--10)~GeV$^2$ for $B$ mesons, with a fixed
ratio $M^2_B/M^2_\rho =5$.  Since for fixed momentum fraction $u$ the
expansion in LC sum rules goes in powers of $1/(uM^2_B)$, we make the
formal replacement \cite{ABS} $M^2_B\to M^2_B/\langle
u\rangle$, where $\langle u\rangle\approx 0.6-0.8$ is the average
momentum fraction calculated by inserting an additional factor $u$
under the integral (separately for each form factor and each value of
$t$), and then taking the interval $M^2_B\approx\,$(4--8)~GeV$^2$, the same as
in the 2pt sum rule for $f_B$.  The scale of condensates and
distribution amplitudes in the sum rules for the form factors is
$\mu^2=m_B^2-m_b^2$.
 
\subsection{Results and Error-Estimates}

Our final results for form factors and spectra are collected in
Figs.~6--8. 
\begin{figure}
\centerline{\epsffile{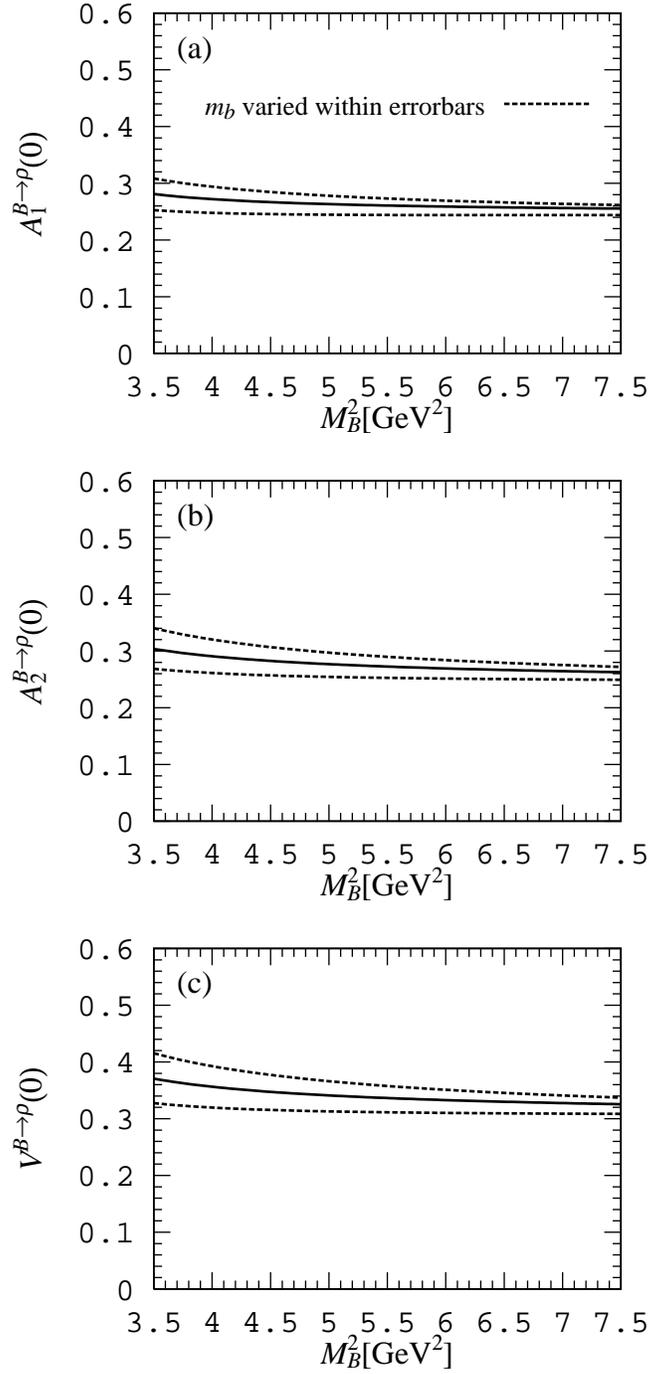}}
\caption[]{Semileptonic $B$ decay form factors at $t=$ as function of
the Borel parameter for central values of the LC sum rule parameters
(solid lines).}
\end{figure}
 First, we display in Fig.~6 the form factors as functions of
the Borel parameter at $t=0$. The solid lines are obtained with 
$m_b = 4.8\,$GeV ($s_0^B = 34\,$GeV$^2$), the dashed lines with $m_b =
4.7\,$GeV ($s_0^B = 35\,$GeV$^2$) and $m_b =4.9\,$GeV ($s_0^B =
33\,$GeV$^2$), respectively. The curves are remarkably flat which
indicates a good accuracy of the sum rules. 
The variation of $a_2^{\parallel,\perp}$ within the specified $\pm
50\%$ range 
has an effect of about the same size as the dependence on $m_b$.
The dependence on the continuum threshold $s_0^B$ is small, provided
the same value is used consistently in the sum rule for $f_B$.
In addition to uncertainties in the sum rule parameters,  
the accuracy of our results is essentially limited by the neglected
higher twist corrections and radiative corrections. 
We have estimated the higher-twist effects for 
$A_1$ in Sec.~3.2 and found them to be approximately $-15$\%. 
This estimate is, however, preliminary and we have not included the
higher-twist correction in our final results in this section.
As for
radiative corrections, we expect them to cancel to some extent when 
 $f_B$ is expressed as 2pt sum rule to $O(\alpha_s^0)$ accuracy. Both
sources of uncertainty can be systematically 
reduced by calculating the
corresponding corrections, which is possible, but beyond the scope of
this paper. Taking everything together, we think that adding an
additional $\pm 15$\% uncertainty to  the above results yields a
fair estimate of the true theoretical error. 

We thus obtain the following values for the form factors at
maximum recoil:
\begin{eqnarray}
A_1^{B\to\rho}(0) & = & 0.27\pm 0.01\pm 0.02\pm 0.02\pm 0.04,\nonumber\\
A_2^{B\to\rho}(0) & = & 0.28 \pm 0.01\pm0.02\pm0.02\pm 0.04,\nonumber\\
V^{B\to\rho}(0) & = & 0.35\pm0.01\pm0.03\pm0.03\pm0.05,
\end{eqnarray}
where the first error comes from the variation in the Borel parameter,
the second from the uncertainty $\pm 0.1\,$GeV in $m_b$, the third
from the uncertainty $\pm 0.1$ in $a_2^{\parallel,\perp}$ and the
forth from the estimated uncertainty due to not included higher twist
and radiative corrections. Note that
the first three errors are correlated between the form factors. 
The results for $A_1(0)$ and $V(0)$ are comparable with those obtained
in \cite{ABS}. In Table~1 we compare our results to  quark
models, adding the errors in quadrature. We have not included the 3pt 
sum rule results \cite{narrdecay,PRD48}, since they suffer from the
deficiencies discussed in Sec.~3. A comparison with lattice
results is difficult, as most of them are obtained at large
$t>14\,$GeV$^2$ and then extrapolated down to $t=0$ using different
assumptions on the functional dependence on $t$ and 
the $b$ quark mass. Only for $A_1$ the
assumed monopole dependence $A_1\sim 1/(m_{B^*}^2-t)$ is compatible
with the scaling law $A_1(0)\sim m_b^{-3/2}$. Using that dependence,
different lattice collaborations have obtained
\begin{equation}
A_1(0) = \left\{ 
\begin{array}{ll}
0.22\pm0.05 & {\rm\ ELC\ \cite{ELC},}\\
0.24\pm0.12 & {\rm\ APE\ \cite{APE},}\\
0.27^{+0.07}_{-0.04} & {\rm UKQCD\ \cite{UKQCD}.}
\end{array}
\right.
\end{equation}
These numbers are quite close to our result.

\begin{table}[t]
\begin{tabular}{lllll}
Reference & $f_+^{B\to \pi}$ & $A_1^{B\to \rho}$ & $A_2^{B\to \rho}$ &
$V^{B\to \rho}$ \\ \hline 
This work & -- & $0.27\pm0.05$ & $0.28\pm0.05$ & $0.35\pm0.07$\\ 
BKR \cite{BKR93} & $0.30$ & -- & -- & -- \\ 
FGM \cite{FGM} & $0.20\pm 0.02$ &
$0.26\pm 0.03$ & $0.31\pm 0.03$ & $0.29\pm 0.03$\\ 
Jaus \cite{jaus} & 0.27 & 0.26 & 0.24 & 0.35 \\
Melikhov \cite{melikhov} & 0.29 & 0.17--0.18 & 0.155 & 0.215 \\ 
WSB \cite{BWS} & 0.33 & 0.28 & 0.28 & 0.33
\end{tabular}
\caption[]{The form factors of the $b\to u$ transitions at $t=0$ in
LC sum rules and quark models. }
\vspace*{0.5cm}
\begin{tabular}{lllll}
Reference & $\Gamma(\bar B^0\to\pi^+e^-\bar\nu)$ & $\Gamma(\bar
B^0\to\rho^+e^-\bar\nu)$ & $\Gamma(\rho)/\Gamma(\pi)$ &
$\Gamma_L/\Gamma_T$ \\ \hline 
This work & -- & $13.5\pm4.0$  & 1.5$\pm$ 0.5 & $0.52\pm0.08$\\ 
BKR \cite{BKR93} & 8.7 & -- & -- & -- \\ 
FGM \cite{FGM} & $3.0\pm 0.6$ & $5.4\pm 1.2$ & -- & $0.5\pm 0.3$\\ 
ISGW2 \cite{ISGWrev} & 9.6 & 14.2 & 1.48 & 0.3 \\ 
Jaus \cite{jaus} & 10.0 & 19.1 & 1.91 & 0.82 \\ 
Melikhov \cite{melikhov} & 7.2 & 9.64 & 1.34 & 1.13\\ 
WSB \cite{BWS} & 7.4 & 26 & 3.5 & 1.34 \\
\end{tabular}
\caption[]{Decay rates of the $b\to u$ transitions in units
$|V_{ub}|^2\,{\rm ps}^{-1}$. $\Gamma_L$ denotes the portion of the
rate with a longitudinal polarized $\rho$ and $\Gamma_T$ with a
transversely polarized $\rho$.}
\end{table}

\begin{figure}
\centerline{\epsffile{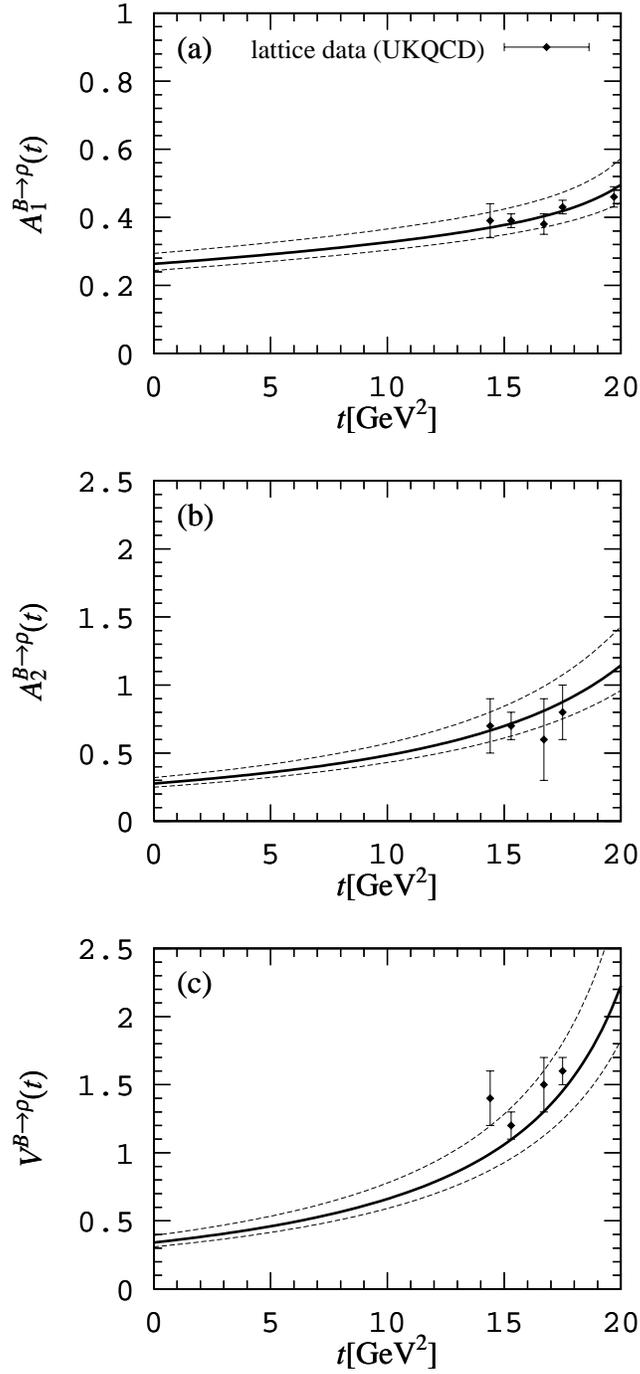}}
\caption[]{Semileptonic B decay form factors as function of $t$ for 
central values of the LC sum rule parameters (solid lines). The dashed
lines give error estimates.}
\end{figure}

Next, we display in Fig.~7 the behaviour of the form factors in $t$
(solid lines) together with error estimates (dashed lines) obtained by
taking extreme values of the parameters: the upper dashed lines
refer to $m_b = 4.7\,$GeV, $M_B^2 = 4\,$GeV$^2$, the lower dashed
lines to $m_b = 4.9\,$GeV, $M_B^2 = 7\,$GeV$^2$. We also show lattice results
from the UKQCD collaboration (diamonds), which are in very good
agreement with our results. The plots indicate clearly that the
accuracy of our results at large $t$ is worse than at small $t$, in
particular for $A_2$ and $V$. However, the contribution of $A_2$ and
$V$ to the experimentally measurable observables, the spectrum in $t$ e.g.,
is kinematically suppressed at large $t$, so that large uncertainties
in that region are not relevant phenomenologically (see also the
discussion below). Figure~7 also shows that $A_1$ is a slowly varying
function of $t$, whereas $A_2$ and $V$ increase more steeply; none of
the form factors can be fitted by a monopole in $t$ as suggested by
the pole-dominance hypothesis. In Ref.~\cite{stech} it was found that
the ratio of form factors takes a simple form in the heavy quark
limit supplemented by some model-assumptions. We find that in the full
range of physical $t$ our ratio $V(t)/A_1(t)$ agrees with the
prediction of \cite{stech} within 4\%, whereas $A_2(t)/A_1(t)$ is by
10\% to 20\% smaller than predicted.

Finally, in Fig.~8, we show the spectra $d\Gamma/dt$ and
$d\Gamma/dE_e$. Fig~8(a) shows the effect mentioned before: although
the uncertainty in the form factors increases with $t$, the
contribution of $A_2$ and $V$ is suppressed and the resulting
uncertainty is dominated by the (smaller) error on $A_1$. The uncertainty
is maximal at $t\approx15\,$GeV$^2$ 
and amounts to $^{+20\%}_{-15\%}$, which
yields a $^{+10\%}_{-8\%}$ accuracy of $|V_{ub}|$ if determined from
that point. Taking into account the additional uncertainties of
unknown higher twist and radiative corrections, we estimate that with
present knowledge $|V_{ub}|$ may be determined from $d\Gamma/dt$ with
a theoretical accuracy of 20\%.
It is conceivable that further calculations may push down this
uncertainty to $\pm 15\%$ on the spectrum, i.e.\ about 8\% on
$|V_{ub}|$, especially if $m_b$  was fixed to better accuracy.
 Fig.~8(b)
also shows that a determination of $|V_{ub}|$ from the 
electron energy spectrum may be more difficult, since it is strongly 
peaked and
the position of the maximum thus may be invisible with presently available
experimental resolution. 

In Ref.~\cite{CLEO} the CLEO collaboration has presented first results
on the branching ratios of $B\to\pi e \nu$ and $B\to\rho e \nu$. Since
the given values  are to a certain extent model-dependent, 
we refrain from extracting any number for
$|V_{ub}|$ from them. This task, we believe, is more appropriate for
our experimental colleagues.

\begin{figure}
\makebox[1cm]{ }
\vspace*{-0.6cm}
\centerline{\epsffile{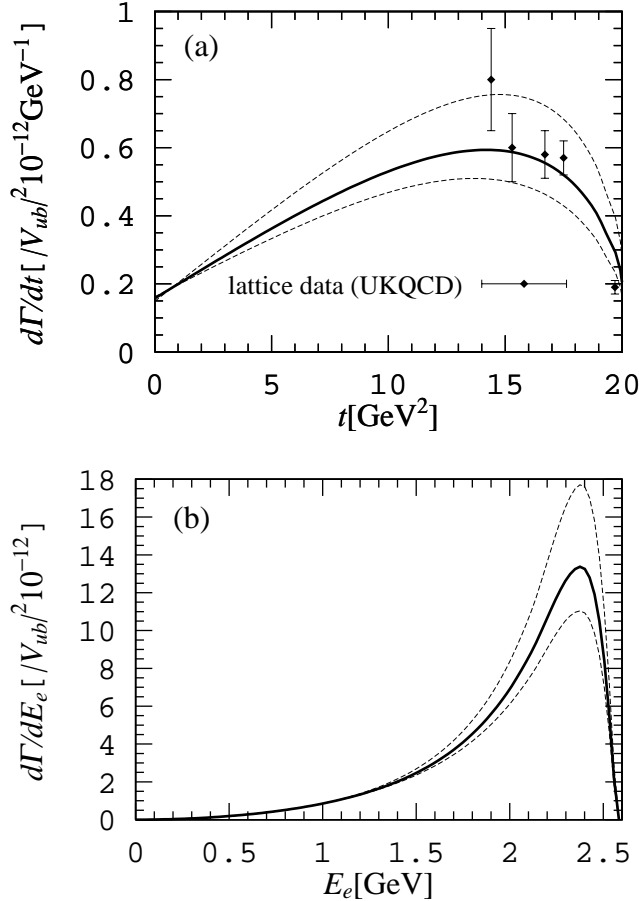}}
\vspace*{-0.5cm}
\caption[]{$B\to\rho e \nu$ decay spectra. (a): spectrum in $t$, (b):
spectrum in the electron energy $E_e$. Solid and dashed lines as in 
Fig.~7.}
\end{figure}

Integrating up the spectra, we find
\begin{equation}
\Gamma(\bar{B}^0\to\rho^+ e^-\bar\nu) = |V_{ub}|^2 (13.5\pm 1.0\pm 1.3\pm
0.6\pm3.6)\, {\rm ps}^{-1} 
\end{equation}
with the same sequence of errors as for the form factors.
In Table~2 we also give ratios of partial decay rates which are
independent of $|V_{ub}|$ and may serve as tests of our
predictions. To get the ratio $\Gamma(\rho)/\Gamma(\pi)$ we have used
the result of \cite{BKR93} obtained by a similar method.


\section{Summary and Conclusions}
\setcounter{equation}{0}

We have given a detailed analysis of existing controversies in
QCD sum rule calculations of semileptonic $B\to\rho e\nu$
form factors, which, as we believe, settles this problem.
Both the  decease of 3pt sum rules, which we have exposed, and the
remedies which we have suggested, apply to all heavy-to-light
transitons and are equally 
relevant e.g.\ for rare radiative decays, where a similar 
discrepancy between LC and 3pt sum rules was found \cite{ABS}.    

We have used the recent reanalysis of $\rho$ meson distribution amplitudes
\cite{rhoWF} to improve and update LC sum rules for the semileptonic 
form factors, including first estimates of higher-twist corrections.
Our final results for the form factors, decay rates  and the spectra
are presented in Tables 1,2 and in Figs.~7,8 together with lattice
data and the results of quark models. We have given a detailed 
analysis of uncertainties of our approach, with the conclusion
that its present accuracy is sufficient for a model-independent 
determination of $|V_{ub}|$ with an error less than 20\%.

The accuracy of our results can be improved, by calculating
radiative corrections to the sum rules and higher-twist
corrections. Both is possible using existing methods and could 
ultimately decrease the uncertainty by a factor two, of order $\sim
10\%$ in   
$|V_{ub}|$. Yet higher accuracy is, however, not feasible within 
the sum rule method.

\bigskip

\noindent{\bf Acknowledgements:}
P.B.\ is grateful to the theory group of NORDITA in Copenhagen for 
its hospitality while this work was finalized.

\bigskip

\noindent{\bf Note added:} When this paper was in writing, the work
\cite{aliev} appeared with a LC sum rule  for $A_2^{B\to K^*}$.
In the SU(3) limit their formula agrees with ours
(except for the $\delta$-function terms related to continuum
subtraction). 


\appendix
\renewcommand{\theequation}{\Alph{section}.\arabic{equation}}
\section*{Appendices}


\section{Continuum Subtraction in LC Sum Rules}
\setcounter{equation}{0}
The ``standard'' procedure, to which we conform in this
paper, consists in approximating the (unknown) physical spectral
function by the perturbative one above some threshold $s_0$, so that
$$
\int_{s_0}^\infty ds\,\frac{\rho^{\rm phys}(s)}{s-p^2} \longrightarrow
\int_{s_0}^\infty ds\,\frac{\rho^{\rm pert}(s)}{s-p^2}.
$$
Thus it is necessary to know the perturbative spectral function explicitly.

In evaluating the correlation function (\ref{eq:cor2pt}) one
encounters terms of type ($q=p_B-p_\rho$, $f(u)$ arbitrary function):
\begin{equation}
I_n = \int_0^1 du\,f(u)\,\frac{1}{[m_b^2-(q+up_\rho)^2]^n}.
\end{equation}
The dispersive representation of $I_1$ is trivial and reads
\begin{equation}
I_1 = -\int_{m_b^2}^\infty ds\, \frac{1}{s-(p_\rho+q)^2}\,\int_0^1 du
\,f(u)\, \delta(us-m_b^2-u\bar u m_\rho^2+t \bar u).
\end{equation}
Putting the upper limit of integration in $s$ to $s_0^B$ simply
introduces a factor $\Theta[c(u,s_0^B)]$ in the integration over
$u$. The function $c$ is defined after Eq.~(\ref{eq:LCV}). For 
higher powers one has either to integrate over $u$ by parts, or  
calculate the spectral function by 
applying consequently two Borel transformations 
$p_B^2\to M^2$ and $1/M^2\to s$, see \cite{doubleB,BB} for
details.
In particular, for $I_2$ we find
\begin{eqnarray}
\rho_{I_2}(s) & = & -\frac{u_*^2}{(m_b^2-t+u_*^2 m_\rho^2)^2}\,\left(
f'(u_*) - \frac{2m_\rho^2 u_* f(u_*)}{m_b^2-t+u_*^2 m_\rho^2}\right),
\end{eqnarray}
where $u_*$ is the solution of $c(u_*,s)=0$ inside the interval
 $0\leq u_*\leq 1$. With this spectral density,
performing the continuum subtraction and the Borel
transformation in $p_B^2$, one obtains after a suitable change of variables
\begin{eqnarray}
\hat{I}_2 - \makebox{continuum} & = & \frac{1}{M^2}\,\int_{u_0}^1
du\,\frac{f(u)}{u^2}\, \exp\left[-\frac{1}{uM^2}\,(m_b^2-\bar u t +
u\bar u m_\rho^2)\right] \nonumber\\
& & {}- \left.\frac{f(u)}{m_b^2-t+um_\rho^2}\, 
\exp\left[-\frac{1}{uM^2}\,(m_b^2-\bar u t + u\bar u m_\rho^2)\right]
\right|_{u_0}^1,
\end{eqnarray}
where $u_0$ is the solution of $c(u_0,s_0^B)=0$ with $0\leq u_0 \leq
1$. Since in our case $f(u)$ 
vanishes at $u=1$, one arrives at the typical
structure $\Theta/(uM^2) + \delta$ that enters (\ref{eq:LCA2}) and
(\ref{eq:LCV}).


\section{3pt Sum Rules with Tensor-Current}
\setcounter{equation}{0}
In this appendix we give the Wilson-coefficients entering the OPE of
$\Sigma^{\mu\nu\alpha}$, Eq.\ (\ref{eq:sigmacorr}). We use the
invariant decomposition
\begin{eqnarray}
\Sigma^{\mu\nu\sigma} & = &
\Sigma_0(g^{\mu\sigma}p_\rho^\nu-g^{\mu\nu}p_\rho^\sigma) +
\Sigma_+(p_B+p_\rho)^\mu(p_B^\nu p_\rho^\sigma-p_B^\sigma
p_\rho^\nu)\nonumber\\
& & {} +i\Sigma_V\left(\epsilon^{\mu\sigma}_{\phantom{\mu\nu}\alpha\beta}
p_B^\alpha p_\rho^\beta p_\rho^\nu - 
\epsilon^{\mu\nu}_{\phantom{\mu\nu}\alpha\beta}p_B^\alpha
p_\rho^\beta p_\rho^\sigma\right)+\dots,
\end{eqnarray}
where $\{\Sigma_0,\,\Sigma_+,\,\Sigma_V\}$ determines the form factors
$\{A_1,\,A_2,\,V\}$. Taking into account perturbation theory, the
quark and the mixed condensate, as well as the four quark condensate
(in vacuum saturation approximation),
the OPE reads:
\begin{equation}
\Sigma = \Sigma^{\rm pert} + \Sigma^{(3)} \langle\bar q q \rangle +
\Sigma^{(5)} \langle \bar q g\sigma G q\rangle + \Sigma^{(6)} \left(-
\frac{16}{9}\right)\alpha_s \pi \langle \bar q q \rangle^2 + \dots
\end{equation}
We give explicit formulas for the Borelized expressions
$\hat{\Sigma}^{(3,5,6)}$ and the double spectral function of
$\Sigma^{\rm pert}$, such that
$$
\Sigma^{\rm pert} = \int ds_b ds_u
\,\frac{\rho^{\rm pert}(s_b,s_u,t)}{(s_b-p_B^2)(s_u-p_\rho^2)} + 
\mbox{subtractions.}
$$
\begin{eqnarray}
\rho^{\rm pert}_0 & = & \frac{3b}{8\pi^2\lambda^{1/2}} +
\frac{3}{8\pi^2\lambda^{3/2}} \,(bT-2s_bs_u)(T-2b),\nonumber\\
\rho^{\rm pert}_+ & = & \frac{3}{4\pi^2\lambda^{3/2}}\,(bT-2s_bs_u) -
\frac{3}{4\pi^2\lambda^{5/2}} \, \{ b(b+2s_u)T^2 - 
3 s_u(b^2+2bs_b+s_bs_u)T\nonumber\\
& & {}+2s_bs_u(b^2+2bs_u+3s_bs_u)\},\nonumber\\
\rho^{\rm pert}_V & = & \frac{3}{2\pi^2\lambda^{3/2}}\,(bT-2s_bs_u) -
\frac{3}{2\pi^2\lambda^{5/2}} \, \{ b^2T^2 - 6bs_bs_u T + 2s_b s_u 
(b^2+3s_bs_u)\},
\end{eqnarray}
with $\lambda = s_b^2+s_u^2+t-2s_bs_u-2s_bt-2s_ut$, 
$b=s_b-m_b^2$ and $T=s_b+s_u-t$. For the nonperturbative terms we
obtain:
\begin{eqnarray}
\hat{\Sigma}^{(3)}_0 & = &
-\frac{m_b}{M_B^2M_\rho^2}\,e^{-m_b^2/M_B^2},
\nonumber\\
\hat{\Sigma}^{(3)}_+ & = & 0,\nonumber\\
\hat{\Sigma}^{(3)}_V & = & 0,\nonumber\\
\hat{\Sigma}^{(5)}_0 & = & \frac{m_b}{M_B^2M_\rho^2}\,e^{-m_b^2/M_B^2}
\left(
  \frac{m_b^2-t}{6M_B^2M_\rho^2} + \frac{m_b^2}{4M_B^4} +
  \frac{1}{6M_\rho^2} - \frac{2}{3M_b^2}\right),\nonumber\\
\hat{\Sigma}^{(5)}_+ & = &
-\frac{m_b}{6M_B^4M_\rho^4}\,e^{-m_b^2/M_B^2},
\nonumber\\
\hat{\Sigma}^{(5)}_V & = &
-\frac{m_b}{3M_B^4M_\rho^4}\,e^{-m_b^2/M_B^2},
\nonumber\\
\hat{\Sigma}^{(6)}_0 & = & \frac{1}{M_B^2M_\rho^2}\,e^{-m_b^2/M_B^2}\left(
  \frac{1}{3M_b^2} - \frac{1}{3M_\rho^2} - \frac{m_b^2}{18M_B^4} -
  \frac{m_b^4}{36M_B^6} - \frac{(m_b^2-t)^2}{18M_B^2M_\rho^4} +
  \frac{2(m_b^2-t)}{9M_\rho^4} \right.\nonumber\\
  & &
  \left.\phantom{\frac{1}{M_B^2M_\rho^2}\,e^{-m_b^2/M_B^2}\left(\right.}
  -\frac{m_b^2(m_b^2-t)}{18M_B^4M_\rho^2}
  + \frac{3m_b^2-2t}{18M_B^2M_\rho^2}\right),\nonumber\\
\hat{\Sigma}^{(6)}_+ & = & \frac{1}{M_B^2M_\rho^2}\,e^{-m_b^2/M_B^2}\left(
  \frac{1}{6M_\rho^4} + \frac{1}{3M_B^2M_\rho^2} +
  \frac{m_b^2}{36M_b^4M_\rho^2} -
  \frac{m_b^2-t}{18M_B^2M_\rho^4}\right),\nonumber\\
\hat{\Sigma}^{(6)}_V & = & \frac{1}{M_B^2M_\rho^2}\,e^{-m_b^2/M_b^2}\left(
  \frac{1}{3M_\rho^4} + \frac{4}{9M_B^2M_\rho^2} +
  \frac{m_b^2}{18M_B^4M_\rho^2} - \frac{m_b^2-t}{9M_B^2M_\rho^4} \right).
\end{eqnarray}


\end{document}